\documentclass[journal=jacsat,manuscript=article]{achemso}

\usepackage[version=3]{mhchem} 

\author{Tzu-chen Liu}
\affiliation{Department of Materials Science and Engineering, Northwestern University}

\author{Adolfo Salgado-Casanova}
\affiliation{Department of Materials Science and Engineering, Northwestern University}

\author{So Yubuchi}
\affiliation{Battery Research Division, Toyota Motor Company}

\author{Bianca Baldassarri}
\affiliation{Department of Materials Science and Engineering, Northwestern University}

\author{Muratahan Aykol}
\affiliation{Energy \& Materials Division, Toyota Research Institue}

\author{Jun Yoshida}
\affiliation{Battery Research Division, Toyota Motor Company}

\author{Hisatsugu Yamasaki}
\affiliation{Advanced Materials Engineering Division, Toyota Motor Company}

\author{Yizhou Zhu}
\affiliation{Department of Materials Science and Engineering, Northwestern University}
\alsoaffiliation{Present Affiliation: Department of Materials Science and Engineering, Westlake University}

\author{Steven B. Torrisi}
\email{steven.torrisi@tri.global}
\affiliation{Energy \& Materials Division, Toyota Research Institue}

\author{Chris Wolverton}
 \email{c-wolverton@northwestern.edu}
\affiliation{Department of Materials Science and Engineering, Northwestern University}

\title
  {Tailored ordering enables high-capacity \\cathode materials}

\begin{document}

\begin{abstract}

Newly designed Li-ion battery cathode materials with high capacity and greater flexibility in chemical composition will be critical for the growing electric vehicles market. 
Cathode structures with cation disorder were once considered suboptimal \cite{whittingham2004lithium}, but recent demonstrations \cite{lee2014unlocking, urban2014configurational, clement2020cation} have highlighted their potential in Li$_{1+x}$M$_{1-x}$O$_{2}$ chemistries with a wide range of metal combinations M. 
By relaxing the strict requirements of maintaining ordered Li diffusion pathways, countless multi-metal compositions in LiMO$_2$ may become viable, aiding the quest for high-capacity cobalt-free cathodes. 
A challenge presented by this freedom in composition space is designing compositions which possess specific, tailored types of both long- and short-range orderings, which can ensure both phase stability \cite{urban2016computational, lun2021cation} and Li diffusion \cite{urban2014configurational, ji2019hidden}. 
However, the combinatorial complexity associated with local cation environments impedes the development of general design guidelines for favorable orderings. 
Here we propose ordering design frameworks from computational ordering descriptors, which in tandem with low-cost heuristics and elemental statistics can be used to simultaneously achieve compositions that possess favorable phase stability as well as configurations amenable to Li diffusion. 
Utilizing this computational framework, validated through multiple successful synthesis and characterization experiments, we not only demonstrate the design of LiCr$_{0.75}$Fe$_{0.25}$O$_2$, showcasing initial charge capacity of 234 mAhg$^{-1}$ and 320 mAhg$^{-1}$ in its 20\% Li-excess variant Li$_{1.2}$Cr$_{0.6}$Fe$_{0.2}$O$_2$, but also present the elemental ordering statistics for 32 elements, informed by one of the most extensive first-principles studies of ordering tendencies known to us.

\end{abstract}

\section{Introduction}

Expanding the scope of metals that could be used in Li-ion battery cathodes could improve capacity, reduce cost, and play a critical role in the worldwide push for vehicle electrification. 
Modern commercial Li-ion battery cathode materials are commonly highly ordered \cite{whittingham2004lithium}, and when Li diffusion primarily relies on ordered pathways \cite{van2013understanding}, inhibited disordering between Li-M becomes a strict requirement in composition selection. 
Nevertheless, recent studies \cite{lee2014unlocking, urban2014configurational, clement2020cation} suggest that the combination of solid solution disorder and reasonable Li diffusivity can coexist in LiMO$_2$, so long as a percolating low-barrier diffusion pathway exists. 
This new finding expands the potential space of orderings and compositions for novel cathode materials design. 
By removing previous assumptions that strict ordering constraints were needed to facilitate electrochemical cycling, a large number of multi-metal mixing permutations for M within LiMO$_2$ became available.
This large number of possible chemistries facilitates the exploration of high-capacity cathodes formulated from more earth-abundant elements, thereby offering potential solutions to availability issues in the global supply chain of cobalt and nickel.

A challenge in designing multi-metal compositions is that not all can achieve a stable single phase and remain electrochemically active. 
This difficulty highlights the need for careful composition design to manage both the stability of desired phases and short-range ordering (SRO) beneficial for Li diffusion in these compounds. 
Recent investigations have circumvented the unfavorable SRO problem by targeting a nearly-ideal random state \cite{lun2021cation, cai2022thermodynamically}, which can be achieved through high-entropy SRO suppression \cite{lun2021cation} or via quenching \cite{cai2022thermodynamically} processes, each of which leverages Li-excess \cite{urban2014configurational} (Li-M ratio $>$ 1) compositions to ensure a percolating low-barrier Li diffusion pathway. 
However, compositional design with favorable SRO can provide ways to further enhance Li diffusion by reducing the Li-content required to form a percolating pathway. 
Furthermore, the focus on fully disordered configurations neglects compositions that might initially be synthesized as single-phase materials (free from competing impurity phases, yet the synthesis temperature is below the disordering temperature) but could be further processed to introduce disorder.
The ability to experimentally understand general ordering trends and design favorable orderings becomes both more challenging but potentially more rewarding as the composition space under consideration and the number of elements grow. 
While computational methodologies \cite{connolly1983density, zunger1990special} have been employed to expedite research on ordering tendencies \cite{urban2016computational, lun2021cation, ji2019hidden}, sampling configurational energetics for disordered states becomes more costly, and the number of local configurations rapidly becomes intractable as the number of elements considered increases.
Due to these difficulties, to our knowledge, no large-scale investigations have yet been conducted on ordering tendencies in multi-metal LiMO$_2$.
In this study, we construct an extensive computational database and design ordering descriptors to identify multi-metal rocksalt-type LiMO$_2$ cathodes exhibiting preferred short- and long-range cation orderings among thousands of compositional possibilities.
The computational framework, supported by experimental synthesis and electrochemical characterization of selected compositions, thereby offers a systematic strategy for materials design and discovery that meets industrial efficiency criteria.
By leveraging insights from our framework, we not only statistically identify elements that enhance targeted rocksalt phase stability and encourage Li-ion diffusion in multi-metal LiMO$_2$ but also demonstrate the design and performance of Li-Cr-Fe-O as a potential high-capacity cathode system that necessitates a modest excess of lithium and incorporates more affordable elements.

\section{Computational Ordering Design Framework}
\begin{figure}
    \centering
    \includegraphics[width=\textwidth]{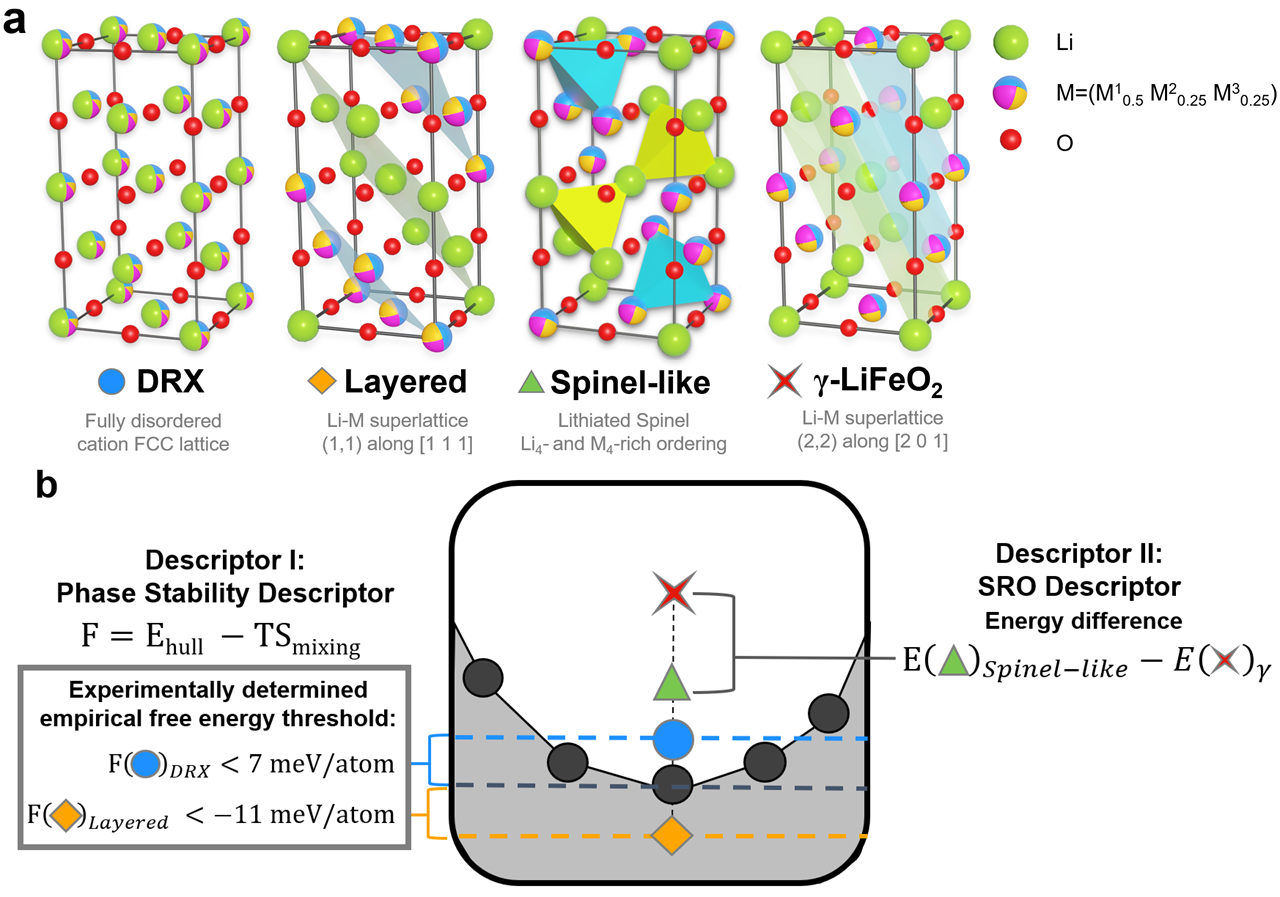}
    \caption{Li-M ordering arrangements in rocksalt-type structures and computational ordering descriptors. \textbf{a.} Common Li-M ordering arrangements\cite{urban2014configurational, wolverton1998cation, hewston1987survey} in rocksalt-type structures with M=M$^1_{0.5}$M$^2_{0.25}$M$^3_{0.25}$ compositions, which serve as the four structural configurations used for deriving our phase stability and SRO descriptors.
    \textbf{b.} Composition stabilities on each arrangement predicted by phase stability descriptors F, compared to OQMD convex hulls (E = 0 reference.) The availability of Li diffusion is predicted by the SRO descriptor, which is defined by the energy difference between Spinel-like and $\gamma$-LiFeO$_2$ Li-M ordering arrangements with M=M$^1_{0.5}$M$^2_{0.25}$M$^3_{0.25}$ sublattice disordering. 
    For more information on the correspondence between the SRO descriptor and Li diffusion, see Figure~\ref{fig:DRX_SROD} and its corresponding section.}
    \label{fig:DRX_structures}
\end{figure}

To develop our ordering design framework, we establish a four-stage process:  
1) First, we constructed a comprehensive density functional theory (DFT) \cite{hohenberg1964inhomogeneous, kohn1965self} database of 24,728 64-atom low-symmetry supercells which captures a large diversity of cation chemistries, 
each in four distinct Li-M cation ordering arrangements \cite{urban2014configurational, wolverton1998cation, hewston1987survey} plotted in Figure~\ref{fig:DRX_structures}a: Layered, Spinel-like, $\gamma$-LiFeO$_2$, and fully cation-disordered rocksalts (DRX), all decorated on the same FCC cation sublattice formed by octahedral interstitial sites of the FCC oxygen sublattice. 
These long-range orderings (LRO) in the rocksalt-type structures are decorated by 6,182 compositions in LiM$^1_{0.5}$M$^2_{0.25}$M$^3_{0.25}$O$_2$, 
LiM$^1_{0.75}$ M$^2_{0.25}$O$_2$ and 
LiM$^1_{0.5}$M$^2_{0.5}$O$_2$, 
where M$^1$, M$^2$ and M$^3$ are chosen from 32 elements.
(See Methods for details on the composition selection process.)
2) Next, we determine the connection between cation composition, ordering behavior, targeted LiMO$_2$ rocksalt phase stability, and Li diffusion. 
We designed simple computational descriptors (Figure~\ref{fig:DRX_structures}b) for both stability and Li diffusion by capturing critical ordering tendencies in multi-metal mixing environments.
3) The correlation between these descriptors and experimental measurements allows us to set empirical thresholds on the values from descriptors for desired properties. 
These thresholds establish guidelines to provide predictive assessment of the entire search space of cathode compositions.
4) The final task is to statistically understand the impact of including different elements on the chance of fulfilling material requirements for rocksalt phase stability and Li diffusion capability, as predicted by the LRO stability and SRO tendency, respectively. 
To summarize, for a given composition, the phase stability descriptor tells us which desirable phases are likely to be stable, and the SRO descriptor predicts the tendencies of Li$_4$ clustering (which are favorable for Li diffusion, as described later and in Figure~\ref{fig:DRX_SROD}a).

\section{Methods}

\subsection{Composition Design}
To comprehensively study ordering tendencies while sampling on a practical compositional space, compositions in the database are generated with the following predefined selection rules for M in LiM$^1_{0.5}$M$^2_{0.25}$M$^3_{0.25}$O$_2$: 
(1) Available oxidation states of M are first determined by the list of ICSD oxidation states provided by pymatgen \cite{ong2013python, belsky2002new}.
(2) Only cations with an octahedral coordination population larger than 8\% in oxides, according to the statistical work reported by Waroquiers et al. \cite{waroquiers2017statistical}, are included. 
(3) Charge balance is maintained. (4) Given oxidation state constraints, cation redox capability of extracting at least 0.5 Li per formula unit is ensured. 
All these rules led to 6,182 compositions after removing duplicates, with their elements placed in the four prototype Special Quasirandom Structures~\cite{zunger1990special}, resulting in 24,728 64-atom supercells in our high-throughput computational database.

\subsection{Special Quasirandom Structures}

Supercells representing Li and M = M$^1_{0.5}$M$^2_{0.25}$M$^3_{0.25}$ configurations in single-phase multi-metal mixing were constructed by the Special Quasirandom Structure (SQS)~\cite{zunger1990special} method.
An SQS is a representative cell which approximates random mixing configurations on a specified lattice of a given size by seeking the cell with cluster site correlations as close as possible to their values in the true random limit, particularly in the first few shells, based on the assumption that correlations between short-range sites typically have a more significant impact on the energy. 
Consequently, the objective function for generating SQS is commonly designed in the following form~\cite{van2013efficient}:
\begin{equation}
\label{eq:SQS-definition}
Q \;=\; -\omega L 
\;+\; \sum_{\alpha}
\Bigl\lvert
  \Gamma_{\alpha}(\sigma^{\mathrm{SQS}})
  \;-\;
  \Gamma_{\alpha}(\sigma^{\mathrm{rnd}})
\Bigr\rvert
\end{equation}
where $\Bigl\lvert
  \Gamma_{\alpha}(\sigma^{\mathrm{SQS}})
  \;-\;
  \Gamma_{\alpha}(\sigma^{\mathrm{rnd}})
\Bigr\rvert$ is the difference in cluster correlations on a mixing (sub)lattice between a generated SQS configuration and the targeted random limit, L is the length that all clusters $\alpha$ with a diameter smaller than L have a perfect correlation match, accompanied by an adjustable weight parameter $\omega$.
While the SQS method is typically applied to perfect-random fully disordered states \cite{urban2016computational, lun2021cation, nyshadham2017computational}, it can be adapted to represent configurations with sublattice disorder or even specific SRO in disordered states \cite{mader1995short, liu2016special, saitta1998structural, van2017software}. 
In this work, 64-atoms whole cation lattice (N) and sublattice (N-1) disordered SQS for four common Li orderings were generated by the Pythonic ICET \cite{aangqvist2019icet} package, cross-checked by an extra set of 64-atom SQS constructed by Alloy Theoretic Automated Toolkit (ATAT) \cite{van2013efficient} to ensure convergence and consistency of correlation calculations between packages. 
These SQS were produced in the element ratio of LiM$^1_{0.5}$M$^2_{0.25}$M$^3_{0.25}$O$_2$, allowing M$^1$=M$^2$ and M$^2$=M$^3$.
In disordered rocksalt structures, cations mix randomly on all cation sites, while for the three other common Li-orderings we study in this work (as seen in Figure~\ref{fig:DRX_structures}a), their SQS maintains the same element ratio, but Li constrained to the corresponding sublattice and M cations mixing on the remaining cation sites. 
These SQS served as prototype structures decorated by the above mentioned 6,182 selected compositions.

\subsection{High-throughput Density-Functional Theory (HT-DFT) Calculations}
HT-DFT calculations for 24,728 64-atoms supercells were executed using the standard Open Quantum Materials Database (OQMD) \cite{saal2013materials, kirklin2015open}  framework, which employs Vienna Ab initio Simulation Package (VASP)\cite{kresse1993ab, kresse1996efficiency, kresse1996efficient} with Projector Augmented Wave (PAW) \cite{blochl1994projector, kresse1999ultrasoft} pseudopotentials and the GGA-PBE \cite{perdew1996generalized} exchange-correlation functionals.
Initial coarse ionic relaxations were performed using OQMD default VASP input settings, converging to 10$^{-3}$ eV for the coarse ionic relaxation loop and 10$^{-4}$ eV for the electronic loop, followed by higher fidelity static calculations with 520 eV cutoff energy for plane wave basis and 8,000 k-points per reciprocal atom (KPPRA). 
Selected candidates in Figure~\ref{fig:DRX_PSD}d,e undergo additional fine ionic relaxations, converging to 10$^{-2}\,\mathrm{eV}/\text{\AA}$ to reach force convergence. 
This was again followed by final static calculations, again with default 520 eV cutoff energy and 8,000 KPPRA, for accurate energy determinations.
The OQMD database is constantly growing with more structures added to pursue more comprehensive and deeper convex hulls but changing over time. 
To maintain consistency, all analyses in this research are based on the convex hull energies acquired in March 2023.

\subsection{Synthesis}

Reported synthesized compounds are prepared using the solid-state method with the following precursors:  
Li$_2$CO$_3$ (Wako, 99.0\%), Cr$_2$O$_3$ (Nacalai Tesque, 98.5\%), Fe$_2$O$_3$ (Kojundo chemical laboratory, 99.9\%), Ga$_2$O$_3$ (Wako, 99.99\%), NiO (Wako, 99.9\%), CuO (Wako, 95.0\%), TiO$_2$ (Wako, 99.0\%), ZrO$_2$ (Wako, 98.0\%).  
Raw materials were mixed in stoichiometric composition, except for Li$_2$CO$_3$, which was set with 3\% excess.  
The mixture of 3 g raw materials and ethanol of 6 mL was put into each 45 mL ball-milling pot.  
The following wet ball-milling process was performed using Pulversettle 7 (FRITSCH) at 300 rpm for 255 minutes, and then the slurry was dried at 120\textdegree{}C under vacuum.  
In the sintering process, powder was placed in a boat-shaped crucible with copper foil wrapped around it.  
Samples were sintered at 1000\textdegree{}C for 12 h in an Ar or air atmosphere.  
The sample was then taken out at 120\textdegree{}C and quickly transferred to a glove box in an Ar atmosphere to minimize air exposure.  
Inside the glove box, the sample was ground with an agate mortar.  
In the additional dry ball-milling process for fine DRX particles, the sintered powder of 1 g was put into each ball-milling pot in an Ar-glove box, which was performed using Pulversettle 7 (FRITSCH) at 600 rpm for total 30 h.

\subsection{Electrochemical Evaluation}

The electrochemical evaluations were performed using the CR-2032 coin cell.  
The weight ratio of active materials/acetylene black/PVdF(KFL\#1120) is 75/13.2/11.8, dispersed and mixed in N-methyl-2-pyrrolidone to form a slurry.  
Next, the slurry was coated on an Al current-collector foil and vacuum-dried at 120\textdegree{}C overnight to produce a positive electrode sheet.  
1 M LiPF$_6$ in EC/DMC/EMC (30/30/40 vol.\%) was used as the electrolyte, and metal Li foil was used as the negative electrode.  
The evaluation was performed at 25\textdegree{}C in a voltage range of 1.5–4.8 V and at a 0.1C rate.

\subsection{XRD Measurement}

Crystal structures were measured using an X-ray diffractometer with Cu-K$\alpha$ radiation using Rigaku Ultima IV in the range of 5–80\textdegree{} with the step 0.005\textdegree{} and scan speed 2\textdegree{} min$^{-1}$ in a dry-air atmosphere.  
The crystalline structure analysis was refined using the computer program PDXL2 (Rigaku Co.).  
The reference Powder Diffraction Files \cite{gates2019powder} (PDF01-089-7118 for Fm-3m and PDF01-072-7839 for R-3m) are from The International Centre for Diffraction Data (ICDD).

\section{Results and Discussion}

\subsection{Phase stability descriptors}

The target rocksalt phases in this research are Layered structures, prevalent in commercial cathodes, and disordered rocksalt (DRX) structures, noted for their promising performance \cite{lee2014unlocking, clement2020cation, lun2021cation}.
The goal is to identify elements that promote the stability of these targeted rocksalt phases for high-capacity cathodes with minimized impurity phases.
The configurational design space can include not only the extremes of completely ordered Layered and fully disordered arrangements, but also states of partial LRO and the transformation between ordered states \cite{ji2020ultrahigh, cai2024situ, shi2021lt, wang2024chemical}, provided these rocksalt phases are electrochemically active.
The transition between the ordered Layered and disordered phases might occur spontaneously through cation migrations during cycling \cite{lee2014unlocking}.
Alternatively, disorder can be manually induced during synthesis via higher furnace temperature or performing additional ball-milling \cite{obrovac1998structure, sato2018metastable}, as demonstrated in the final high-capacity cathode design  section. 
We anticipate future generalization of these descriptors to other beneficial LRO, including Spinel-like ordering (which is generally slightly higher in energy than Layered ordering, as shown later in Figure~\ref{fig:DRX_LvsSP}), and the precise control of the partial disordering level.

The stability of phases was investigated by the energetic competition between various Li orderings in single-phase multi-metal mixing structures and the ground-state (combination of) structures in the corresponding phase space. 
For each Li ordering type, the phase stability was determined using an approximate free energy F = E$_{\mathrm{hull}}$ - TS$_{\mathrm{mixing}}$, where here T = 1273K is the standard synthesis temperature adopted in experimental results described later, E$_{\mathrm{hull}}$ is the difference in energy between the multi-metal mixing compound and the ground state energy from the OQMD, and S$_{\mathrm{mixing}}$ is the ideal configurational entropy from cation sublattice mixing allowed in each ordering type.  
The use of an ideal mixing entropy is where the main approximation to the free energy enters.
We computed F for all 6,182 compositions for each of the four Li-M orderings (DRX, Layered, Spinel, and $\gamma$-LiFeO$_2$) and studied their distributions (Figure~\ref{fig:DRX_PSD}a,b,c). 
A lower value of F indicates increased stability with respect to competing phases, and hence an increased likelihood for experimental synthesis.  
We note that the approximate free energy F of the designed structures, used as our descriptor, inherently excludes kinetics and metastability of potential phases that influence realistic synthesizability. 
Nevertheless, the agreement between our computational predictions and subsequent experimental samples underscores the fundamental role of thermodynamic stability in predicting the synthesized phases. 
As we show next, by comparing our DFT free energy predictions with experimental synthesis attempts, we are able to determine a threshold value for  F that predicts whether a compound can be experimentally synthesized in targeted rocksalt phases. 

\begin{figure}
    \centering
    \includegraphics[width=1\textwidth]{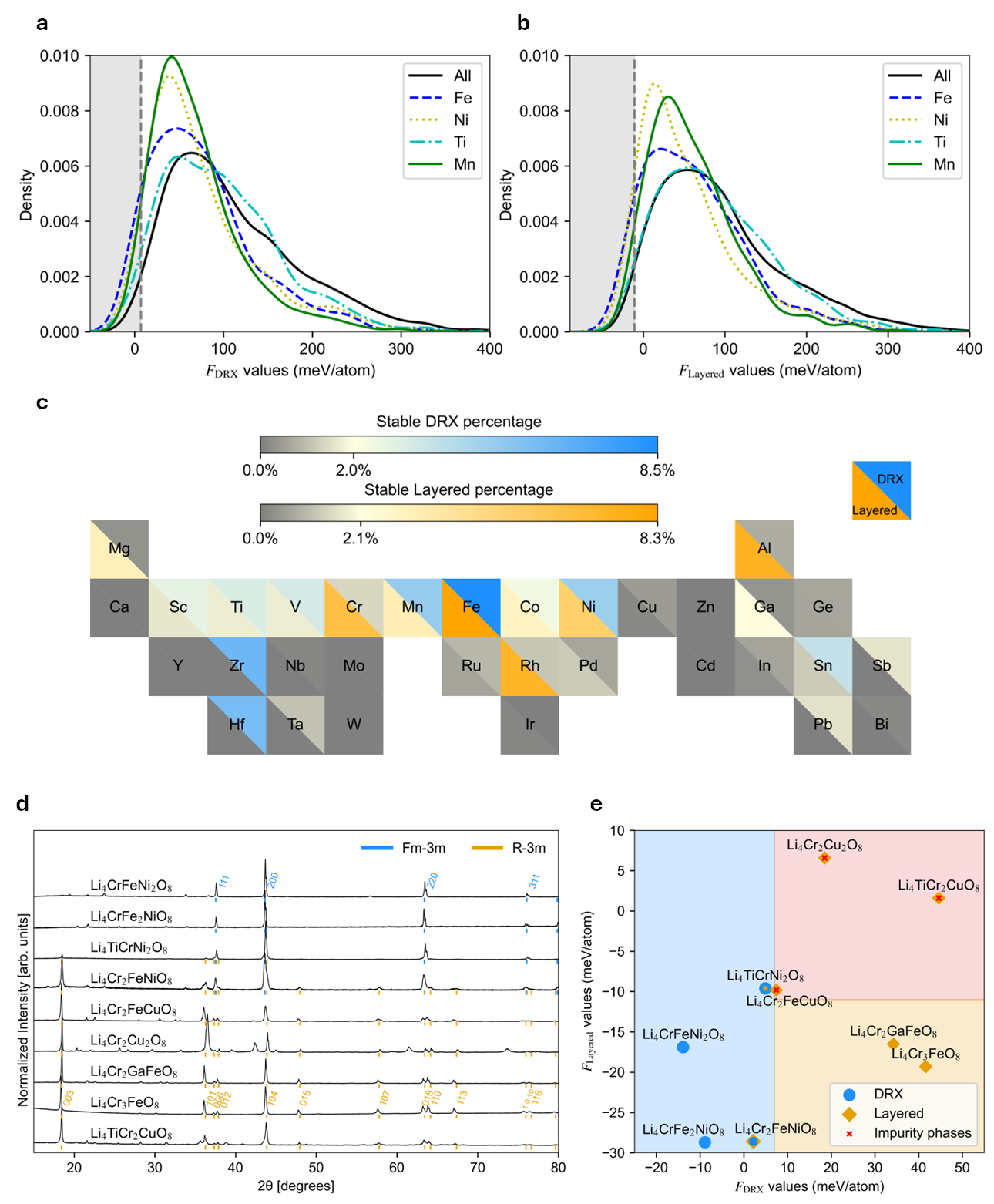}
    \caption{Phase stability descriptors F and elemental statistics for stable DRX and Layered phases. 
    {a.} F$_{\mathrm{DRX}}$ and \textbf{b.} F$_{\mathrm{Layered}}$ distributions of 6,182 compositions from the database. 
    The black curve represents the F distribution of all compositions.
     (Continued on next page.)} 
    \label{fig:DRX_PSD}
\end{figure}

\begin{figure}
  \ContinuedFloat
  \caption{(cont.) 
    Individual curves for each element depict distributions calculated from compositions containing that specific element, demonstrated with examples of Fe, Ni, Ti, and Mn. 
    Gray dotted vertical lines are threshold values of F$_{\mathrm{DRX}}$ and F$_{\mathrm{Layered}}$ empirically derived from synthesized compounds shown in subfigures d and e, with the compositions in the shaded area classified as stable in DRX/Layered when synthesized at 1273K. Additional details on F value distributions for all 32 elements are provided in the Supporting Information.
  \textbf{c.} Periodic-table-style heat map for the percentages classified as stable DRX (upper right triangle) and Layered (bottom left triangle) phases, illustrating the influence of each element on stabilizing target phases. The color bar is configured to drastically change around the value of stable percentage from “All” compositions, 2.0\% for DRX and 2.1\% for Layered phases.
  \textbf{d.} XRD spectra of synthesized compositions. Peaks corresponding to the DRX phases (space group Fm-3m) are labeled with blue lines and plane indices displayed nearby. Similarly, peaks associated with Layered phases (space group R-3m) are indicated with orange lines and their corresponding plane indices. Other peaks identified as impurity phases are not labeled to maintain clarity. 
  \textbf{e.} F$_{\mathrm{DRX}}$ and F$_{\mathrm{Layered}}$ values of synthesized compositions, with icons indicating their experimentally observed phases as determined by XRD analysis in subfigure d. 
  Blue circles indicate the DRX phase, orange diamonds represent the Layered phase, and red crosses represent impurity phases. 
  The compositions are categorized into distinct regions based on empirically derived threshold values, 7 meV/atom for F$_{\mathrm{DRX}}$ and -11 meV/atom for F$_{\mathrm{Layered}}$.
  The F values in this plot are calculated using higher-fidelity E$_{\mathrm{hull}}$ examined by additional fine ionic relaxations specifically for candidate materials to further enhance structural optimizations. 
  More details of this fine ionic relaxation examination on E$_{\mathrm{hull}}$ are discussed in the Supporting Information.}
  \label{fig:PSD-cont}
\end{figure}

To determine the classification threshold for descriptor values required to stabilize target phases, we performed a series of synthesis and X-ray diffraction (XRD) experiments for compositions displayed in Figure~\ref{fig:DRX_PSD}d, totaling nine synthesized compositions. 
These compositions were chosen from a set with relatively low values of F$_{\mathrm{DRX}}$ and F$_{\mathrm{Layered}}$ (and high F$_{\gamma\textrm{-LiFeO}_2}$) to explore the transition from stable DRX or Layered orderings to the emergence of impurity phases.  
Among synthesized compositions, all compounds showing the existence of the DRX phase exhibit lower F$_{\mathrm{DRX}}$ values.
Observing the synthesis of the DRX phase allows us to establish an empirical threshold around 7 meV/atom for classifying compositions, with F$_{\mathrm{DRX}}$ below this value forming stable DRX. 
As for the remaining compositions whose DRX phase still cannot be stabilized with T=1273 K, their F$_{\mathrm{Layered}}$ values can be used to predict the stability of single-phase Layered structures without significant impurity phases. 
It is noteworthy that F$_{\mathrm{Spinel\text{-}like}}$ values are closely correlated with and generally slightly higher than F$_{\mathrm{Layered}}$ values, a trend observed within the nine compositions we synthesized and also, on average, across the entire computational database as shown in Figure~\ref{fig:DRX_LvsSP}.
This observation agrees with the fact that the Layered phase is more common than the Spinel-like phase in synthesized LiMO$_2$ \cite{hewston1987survey, wu1998size, urban2014configurational}, despite both orderings exhibiting the same pair and three-body multi-site cluster correlations \cite{wolverton1998cation}.
The threshold values for DRX and Layered phase stability are determined by comparison with experimental synthesis data as 7 and -11 meV/atom, respectively, to effectively classify all synthesized compositions, as demonstrated in Figure~\ref{fig:DRX_PSD}e. 
The accuracy of the threshold can be improved systematically by performing more experiments guided by computational predictions near the threshold, aiming for a high level of self-consistency for classifications. 
For several selected compositions, we also synthesized variants with a 20\% excess of Li and observed phases nearly identical to those of the non-Li-excess compositions, as detailed in Supporting Information Figure~\ref{fig:DRX_Li12check}.
These observations suggest an additional compositional flexibility of the Li-excess level, Li$_{1+x}$M$_{1-x}$O$_{2}$, in the design of targeted phases. 
While we find that predictions for compositions with up to 20\% Li-excess can be inferred from descriptors based on stoichiometric structures, extending the current framework to higher levels of Li-excess can also be achieved by introducing the desired amount of excess Li into the prototypes shown in Figure~\ref{fig:DRX_structures}a.

\begin{figure}
    \includegraphics[width=\textwidth]{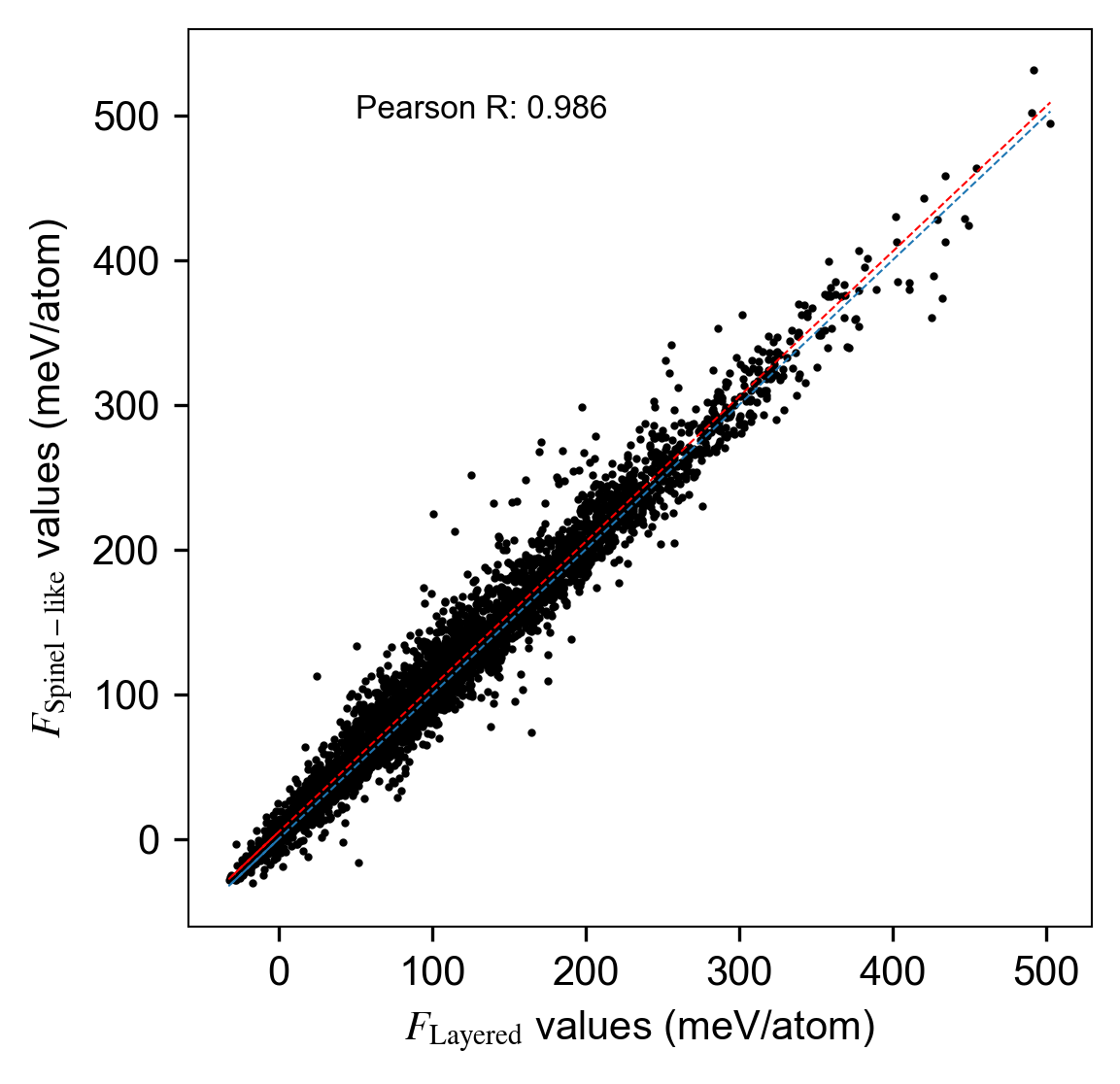}
    \caption{Scattering plots of F$_{\mathrm{Layered}}$ and F$_{\mathrm{Spinel\text{-}like}}$ values of all 6,182 compositions. 
    The lower blue dotted line is the F$_{\mathrm{Layered}}$ = F$_{\mathrm{Spinel\text{-}like}}$ and the upper red dotted line is from linear regression. In the cluster description, Layered and Spinel-like orderings exhibit identical pair and three-body correlations, diverging only starting from four-body correlations \cite{wolverton1998cation}.
    With this similarity, it is expected that the energies of these two phases are highly correlated. 
    Still, 68.4\% of the compositions are above the blue dot line, and the average F$_{\mathrm{Layered}}$ values are lower than F$_{\mathrm{Spinel\text{-}like}}$ by 5.1 meV/atom. 
    This energy difference is closely related to four-body interactions in cluster expansions.}
    \label{fig:DRX_LvsSP}
\end{figure}

The final task in this section is to examine the impact of each element in obtaining the targeted rocksalt phases without impurity phases upon experimental synthesis. 
We analyzed the distributions of descriptor values across the entire computational database of 6,182 compositions, evaluating the percentages of compositions passing DRX and Layered empirical thresholds, respectively, within subsets containing each metal species (Figure~\ref{fig:DRX_PSD}c). 
Elements such as Sc, Ti, Zr, Hf, V, Mn, Fe, Co, Ni and Sn for DRX structures or Mg, Cr, Mn, Fe, Co, Rh, Ni, Al and Ga for Layered structures are statistically more likely to be beneficial for the stability inside multi-metal mixing environments. 
Still, with mixing of up to 4 cations, distributions of F show considerable variability. 
While adding specific elements has some statistical benefit toward the likelihood of synthesis, it is not true that any composition containing those elements can guarantee completely ordered Layered or fully disordered phases. 
However, the combinations of elements mentioned above are likely to have a higher chance of success in achieving the desired phases, and the most promising compositions can be quickly identified using the descriptors and phase stability thresholds. 
Additionally, the elemental analysis above is not significantly sensitive to the choices of threshold values, as demonstrated in the Supporting Information.
We note that our work is distinct from previous DRX-focused research \cite{lun2021cation} in terms of the anion chemistries considered (fluorinated vs. pure oxides), the cation compositions (high-Li-excess vs. stoichiometric), and the targeted ordering (high-entropy random vs. favorable orderings).

\subsection{Li diffusion and SRO descriptors}

Facile Li diffusion is necessary for high-capacity cathode materials.
One mechanism proposed by Lee et al. \cite{lee2014unlocking} states that when disordering occurs, Li diffusion in rocksalt-type oxides can be aided by the formation of percolating channels of sites with low diffusion barriers.  
These channels are constructed by local “Li$_4$ clustering”, a cluster of four Li atoms positioned on four octahedral cation sites surrounding the tetrahedral site on the Li diffusion pathway, as contrasted with arrangements where there are one or more transition metals on the cation sites. 
This “0-TM” \cite{lee2014unlocking, urban2014configurational} local configuration avoids the strong electrostatic repulsion between the diffusing Li$^+$ and high-valence TM cations, yielding lower barriers, and hence the major Li diffusion channels after disordering. 
Ji et al. \cite{ji2019hidden} has demonstrated that SRO preference toward more Li-M mixing in disordered configurations decreases the availability of Li$_4$ clustering. 
However, accurately simulating SRO in disordered systems is not feasible for the over 6,000 compositions in this study. 
To identify compositions and elements that foster targeted SRO exhibiting more Li$_4$ clustering (Figure~\ref{fig:DRX_SROD}a) after disordering, we designed an SRO descriptor as the computational energy difference between Spinel-like and $\gamma$-LiFeO$_2$ type Li-M ordering arrangements with M=M$^1_{0.5}$M$^2_{0.25}$M$^3_{0.25}$ mixing in a single-phase low-symmetry environment: SRO descriptor = E$_{\mathrm{Spinel\text{-}like}}$ - E$_{\gamma\textrm{-LiFeO}_2}$.
As demonstrated in Figure~\ref{fig:DRX_SROD}a, the Spinel-like and $\gamma$-LiFeO$_2$ structures contain the largest distinction in the ratios of Li$_4$ and Li$_2$M$_2$ local environments (the most Li-M mixed local configuration) in ordered LiMO$_2$ rocksalt-type structures. 
The value of this descriptor is therefore related to the energy cost required to switch atom positions from Li$_2$M$_2$ to Li$_4$ + M$_4$ clustering configurations in low-symmetry environments. 
Negative values mean more energetically favorable Li$_4$ clustering, hinting at less Li-M mixing and more Li$_4$ clustering SRO tendency, which would induce a greater likelihood of a 0-TM percolating network and thus better Li diffusion. 
We note that the SRO descriptor neglects potential strong ordering tendencies within the M sublattice, operating under the hypothesis that such interactions are unlikely to become significant enough to outweigh the Li-M interactions captured by the SRO descriptor in Li$_4$ clustering tendency. 

\begin{figure}
    \centering
    \includegraphics[width=1\textwidth]{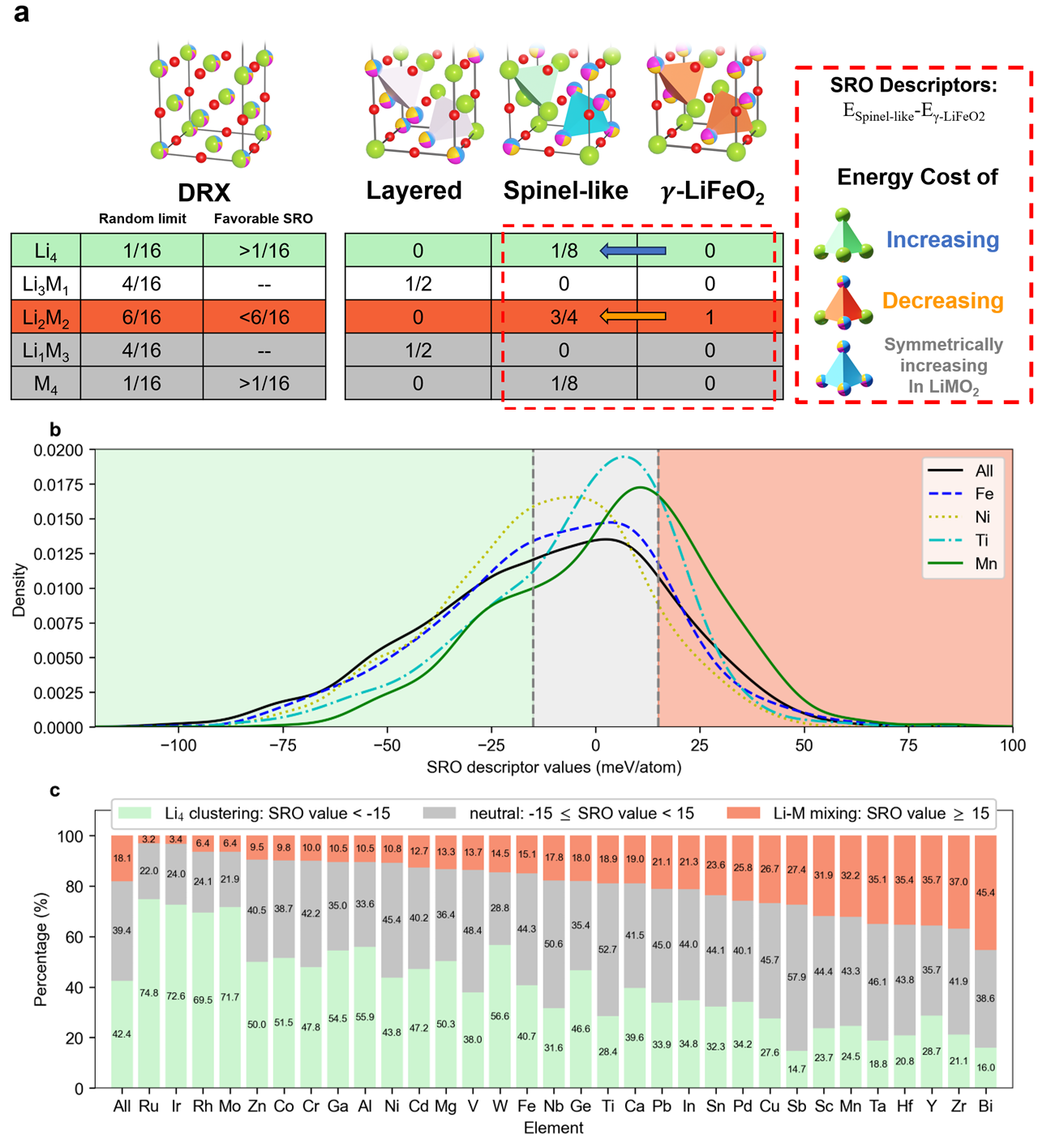}
    \caption{SRO descriptor and value distributions.  
    \textbf{a.} Distributions of local $\mathrm{Li}_{x}\mathrm{M}_{4-x}$ configurations in four Li-M arrangements.
    (Continued on next page.)
     }
    \label{fig:DRX_SROD}
\end{figure}

\begin{figure}
  \ContinuedFloat
  \caption{(cont.) Unlike the binomial distribution from perfect-random DRX, the other three Li-M orderings each feature distinct $\mathrm{Li}_{x}\mathrm{M}_{4-x}$
  local configurations. The SRO descriptor is designed by comparing Spinel-like and $\gamma$-LiFeO$_2$ Li orderings in low-symmetry cation-mixing conditions to capture the energy difference between Li$_4$-rich vs Li$_2$M$_2$-rich cation arrangements. 
  The distinction of $\mathrm{Li}_{x}\mathrm{M}_{4-x}$ configurations between Spinel-like and $\gamma$-LiFeO$_2$ can be conceptually considered as the transition of $\frac{1}{4}$ Li$_2$M$_2$ to $\frac{1}{8}$ Li$_4$ and M$_4$, and the corresponding energy difference showing the tendency for mixing and clustering.
  \textbf{b.} Smeared distributions of SRO descriptor values from the computational database. The black curve represents the distribution of values from all compositions.
  Individual curves for each element illustrate the distributions from only compositions including that specific element, demonstrated with Fe, Ni, Ti, and Mn examples. 
  The gray dot vertical lines are the thresholds to categorize compositions into three classes: “Li-M mixing” SRO (descriptor) value $>$ 15 meV/atom, “neutral” -15 $<$ SRO (descriptor) value $<$ 15 meV/atom, and “Li$_4$ clustering” SRO (descriptor) value $<$ -15 meV/atom. 
  \textbf{c.} The percentages of compositions containing each specific element in each class. 
  The plot demonstrates elemental energetic tendencies toward Li$_4$ clustering or Li-M mixing local configurations, with the sorted elements on the right introducing more compositions that exhibit stronger Li-M mixing to interrupt Li diffusion.
    }
\label{fig:SROD-cont}
\end{figure}

We first validate the SRO descriptor by reproducing observations from previous study \cite{ji2019hidden} of Li$_2$MnTiO$_4$ and Li$_2$MnZrO$_4$. 
We calculated the descriptor value of Li$_2$MnZrO$_4$ as 41 meV/atom, reflecting a strong Li-M mixing SRO tendency that restricts the Li diffusion, consistent with the relatively slow diffusion in (a 20\% Li-excess variant of) this compound \cite{ji2019hidden}. 
The SRO descriptor value of Li$_2$MnTiO$_4$ is 11 meV/atom, much lower than the Zr counterpart, aligning with its ordering nature closer to the random limit.  
However, the descriptor remains positive, which reflects previous observations that Li$_4$ populations in both compositions are lower than the random limit. 
Utilizing this simple SRO descriptor, we capture the primary Li$_4$ clustering SRO tendency critical for 0-TM percolation. 
Next, we further validated our descriptor against other experimentally synthesized compositions (Table \ref{tab:DRX_SRO}) and found that most reported DRX compositions have SRO descriptor values either close to zero or negative. 
Quantitative SRO measurements, though crucial as direct evidence of SRO, are challenging to perform and are rarely reported in the cited literature. 
Nonetheless, the observed connection between active electrochemical performance, the target quantity that SRO is designed to support, and our SRO descriptor provides additional, albeit indirect, qualitative validation that can be further leveraged.
We noted that this qualitative assessment might be biased against negative results, as electrochemically inactive results are rarely reported.

\begin{table}
  \centering
  \caption{\label{tab:DRX_SRO}SRO descriptors of reported electrochemically active compositions, except for Li$_{1.2}$ Mn$_{0.4}$ Zr$_{0.4}$ O$_2$, which exhibits unfavorable SRO consistent with the SRO descriptor and facilitates the classification of SRO type in Figure~\ref{fig:DRX_SROD}. We note the extreme case of Li$_2$Ni$_{0.333}$Ru$_{0.666}$O$_3$ and Li$_{1.211}$Mo$_{0.467}$Cr$_{0.3}$O$_2$, which exhibit such strong clustering tendency that they form LRO (Spinel-like and Layered type, respectively) rather than SRO in their as-synthesized phases. }
  {
  \renewcommand{\arraystretch}{1.2}
  \begin{tabular}{c c c c}
    \hline
    \shortstack[c]{Composition in\\computational dataset}
      & \shortstack[c]{SRO descriptor\\(meV/atom)}
      & \shortstack[c]{Synthesized\\Composition}
      & Reference \\
    \hline
    Li$_2$TiMnO$_4$        &   11 & Li$_{1.2}$Mn$_{0.4}$Ti$_{0.4}$O$_2$            & \cite{ji2019hidden}  \\ \hline
    Li$_2$ZrMnO$_4$        &   41 & Li$_{1.2}$Mn$_{0.4}$Zr$_{0.4}$O$_2$            & \cite{ji2019hidden}  \\ \hline
    Li$_2$MnVO$_4$         &   10 & Li$_2$MnVO$_4$                                & \cite{cambaz2019design} \\ \hline
    Li$_2$FeVO$_4$         &  -15 & Li$_2$FeVO$_4$                                & \cite{cambaz2019design} \\ \hline
    Li$_2$CoVO$_4$         &  -11 & Li$_2$CoVO$_4$                                & \cite{cambaz2019design} \\ \hline
    Li$_2$TiNiO$_4$        &   -4 & Li$_{2.1}$NiTiO$_{3.77}$                      & \cite{prabaharan2004li2nitio4} \\ \hline
    Li$_2$TiCoO$_4$        &   -4 & Li$_2$CoTiO$_4$                              & \cite{yang2012cation} \\ \hline
    Li$_2$TiVO$_4$         &   -4 & Li$_{2-x}$VTiO$_4$                            & \cite{dominko2011electrochemical} \\ \hline
    Li$_2$TiFeO$_4$        &    9 & \shortstack[l]{Li$_2$TiFeO$_4$ \\ Li$_{1.2}$Ti$_{0.4}$Fe$_{0.4}$O$_2$} &  \shortstack[l]{\cite{kuzma2009electrochemical}\\ \cite{yamamoto2019charge}} \\ \hline
    Li$_4$TiVFe$_2$O$_8$    &   -7 & Li$_2$Fe$_{0.5}$Ti$_{0.5}$O$_4$                & \cite{chen2016identifying} \\ \hline
    Li$_2$MnNbO$_4$        &    7 & Li$_{1.3}$Nb$_{0.3}$Mn$_{0.4}$O$_2$            & \cite{yabuuchi2016origin} \\ \hline
    Li$_2$NbFeO$_4$        &    1 & Li$_{1.3}$Nb$_{0.3}$Fe$_{0.4}$O$_2$            & \cite{yabuuchi2016origin} \\ \hline
    Li$_2$NbVO$_4$         &   -5 & Li$_{1.3}$Nb$_{0.3}$V$_{0.4}$O$_2$             & \cite{yabuuchi2016origin} \\ \hline
    Li$_2$NbNiO$_4$        &   -6 & Li$_{1.2}$Ni$_{0.4}$Nb$_{0.4}$O$_2$            & \cite{lin2021fundamental} \\ \hline
\shortstack[l]{Li$_4$Ti$_2$NbNiO$_8$\\ Li$_4$TiNbNi$_2$O$_8$}
  & \shortstack[c]{-1\\ -3}
  & Li$_{1.2}$Ni$_{0.3}$Ti$_{0.3}$Nb$_{0.2}$O$_2$
  & \cite{yu2019synthesis} \\ \hline

\shortstack[l]{Li$_4$Ti$_2$NiMoO$_8$\\ Li$_4$TiNi$_2$MoO$_8$}
  & \shortstack[c]{-34\\ -20}
  & Li$_{1.2}$Ni$_{0.333}$Ti$_{0.333}$Mo$_{0.133}$O$_2$
  & \cite{kwon2020impact} \\ \hline

\shortstack[l]{Li$_2$NiRuO$_4$\\ Li$_4$NiRu$_3$O$_8$}
  & \shortstack[c]{-35\\ -53}
  & Li$_2$Ni$_{0.333}$Ru$_{0.666}$O$_3$
  & \cite{li2019new} \\ \hline
    Li$_2$TaNiO$_4$        &  -11 & Li$_{1.3}$Ni$_{0.27}$Ta$_{0.43}$O$_2$          & \cite{jacquet2019charge} \\ \hline
    Li$_2$MoCrO$_4$        &  -56 & Li$_{1.211}$Mo$_{0.467}$Cr$_{0.3}$O$_2$        & \cite{lee2014unlocking}  \\ \hline
    Li$_2$TaMnO$_4$        &    6 & Li$_{1.3}$Ta$_{0.3}$Mn$_{0.4}$O$_2$            & \cite{kan2019evolution} \\ \hline
  \end{tabular}
  }
  
\end{table}

Considering the approximate upper range of the SRO descriptor value for reported electrochemically active compositions, we define an approximate threshold of 15 meV/atom and screen candidate compositions to identify those with SRO favorable for Li diffusion. 
Given that most reported DRX compositions have SRO descriptor values close to zero, we classified the SRO descriptor value into three categories (Figure~\ref{fig:DRX_SROD}b): 
``Li-M mixing'' (SRO descriptor value $>$ 15 meV/atom), ``neutral'' (15 $>$ SRO descriptor value $>$ -15 meV/atom), and ``Li$_4$ clustering'' (SRO descriptor value $<$ -15 meV/atom), which can be qualitatively considered as “Poor”, “Moderate”, and “Good” for Li diffusion, respectively.
An overview of the SRO descriptor distributions over all compositions is given in Figure~\ref{fig:DRX_SROD}c.
Nearly 40\% of compositions exhibit values within the narrow neutral region.
While 38\% of compositions have positive values, only 18\% fall into the “Li-M mixing” class, indicating limited Li diffusion. 
Our design guideline for encouraging Li diffusion in newly proposed compositions is to avoid compositions with SRO descriptor values exceeding 15 meV/atom, as a large positive value indicates a stronger Li-M mixing tendency. 
Such compositions might need a high Li-excess level to obtain a percolation threshold \cite{urban2014configurational}.
Next, 42\% of compositions are classified as “Li$_4$ clustering”, energetically preferring the Li$_4$ clustering environment over Li-M mixing, which lowers the Li-excess level needed to activate Li diffusion.

Once again, we explore the impact of each element on SRO and Li diffusion by comparing the statistics of SRO descriptor values (Figure~\ref{fig:DRX_SROD}c) across all compositions versus those containing each element. 
On average, elements such as Bi, Zr, and a few more elements with higher SRO values (right-hand side of Figure~\ref{fig:DRX_SROD}c) introduce more Li-M mixing. 
These elements should be carefully examined, and their content decreased or excluded to protect Li diffusion.
To reduce Li-M mixing by obtaining SRO close to the ``neutral" in solid solutions, compositions with elements such as Ti, Nb, and V have around 50\% of SRO descriptor values in the “neutral” range, and their median and mean values are also close to zero.
This tendency might be attributed to their adaptability to various local environments \cite{urban2017electronic}, consistent with the d$_0$ TM cation guideline in DRX.
Elements such as Ru and Ir with more negative SRO values (left-hand side of Figure~\ref{fig:DRX_SROD}c) might bring more Li$_4$ clustering, hence require lower amounts of Li-excess for diffusion, enable more redox capacity simply from the TM and minimize the risk of irreversible oxygen redox for better cyclability \cite{lee2021determining}.
The elemental ordering tendency ranking in Figure \ref{fig:DRX_SROD}c is again not sensitive to threshold variations, as discussed in Supporting Information. 
We note that, while individual elements can statistically contribute toward favorable orderings, it is incorrect to assume that any composition containing those elements will guarantee favorable SRO, as the simplified SRO descriptor does not fully account for all potential ordering interactions on the M sublattice, which may be significant in specific cases.

\section{High-capacity cathode design}

We demonstrate the effectiveness of these computational descriptors and elemental statistics toward favorable ordering through the case of the Li-Cr-Fe-O cathode. 
As shown in Figure~\ref{fig:DRX_PSD}d,e, LiCr$_{0.75}$Fe$_{0.25}$O$_2$, LiCr$_{0.5}$Ga$_{0.25}$Fe$_{0.25}$O$_2$, LiTi$_{0.25}$Cr$_{0.25}$Ni$_{0.5}$O$_2$, and different ratios of (Cr, Fe, Ni) combinations are successfully synthesized as Layered or DRX phases with negligible impurity phases. 
Of the candidates considered, LiCr$_{0.75}$Fe$_{0.25}$O$_2$, which displayed the most negative SRO descriptor value = -32 meV/atom, is chosen for electrochemical testing on both LiCr$_{0.75}$Fe$_{0.25}$O$_2$ and its 20\% Li-excess variant Li$_{1.2}$Cr$_{0.6}$Fe$_{0.2}$O$_2$.

\begin{figure}
    \centering
    \includegraphics[width=1\textwidth]{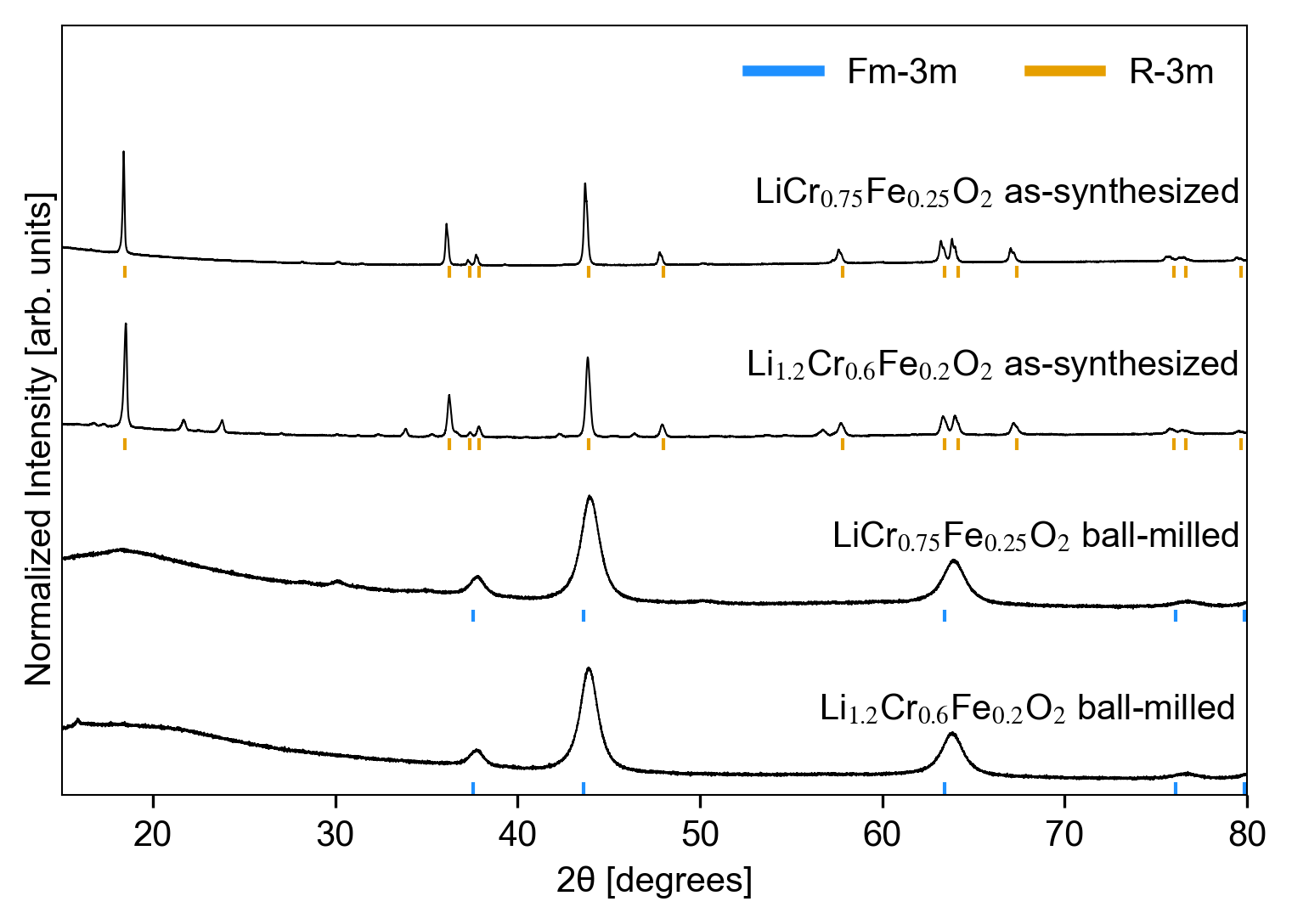}
    \caption{XRD spectra of as-synthesized (Layered phases) and ball-milled (DRX phases) LiCr$_{0.75}$Fe$_{0.25}$O$_2$ and Li$_{1.2}$Cr$_{0.6}$Fe$_{0.2}$O$_2$.
     }
    \label{fig:DRX_CrFe12}
\end{figure}

\begin{figure}
    \centering
    \includegraphics[width=0.8\textwidth]{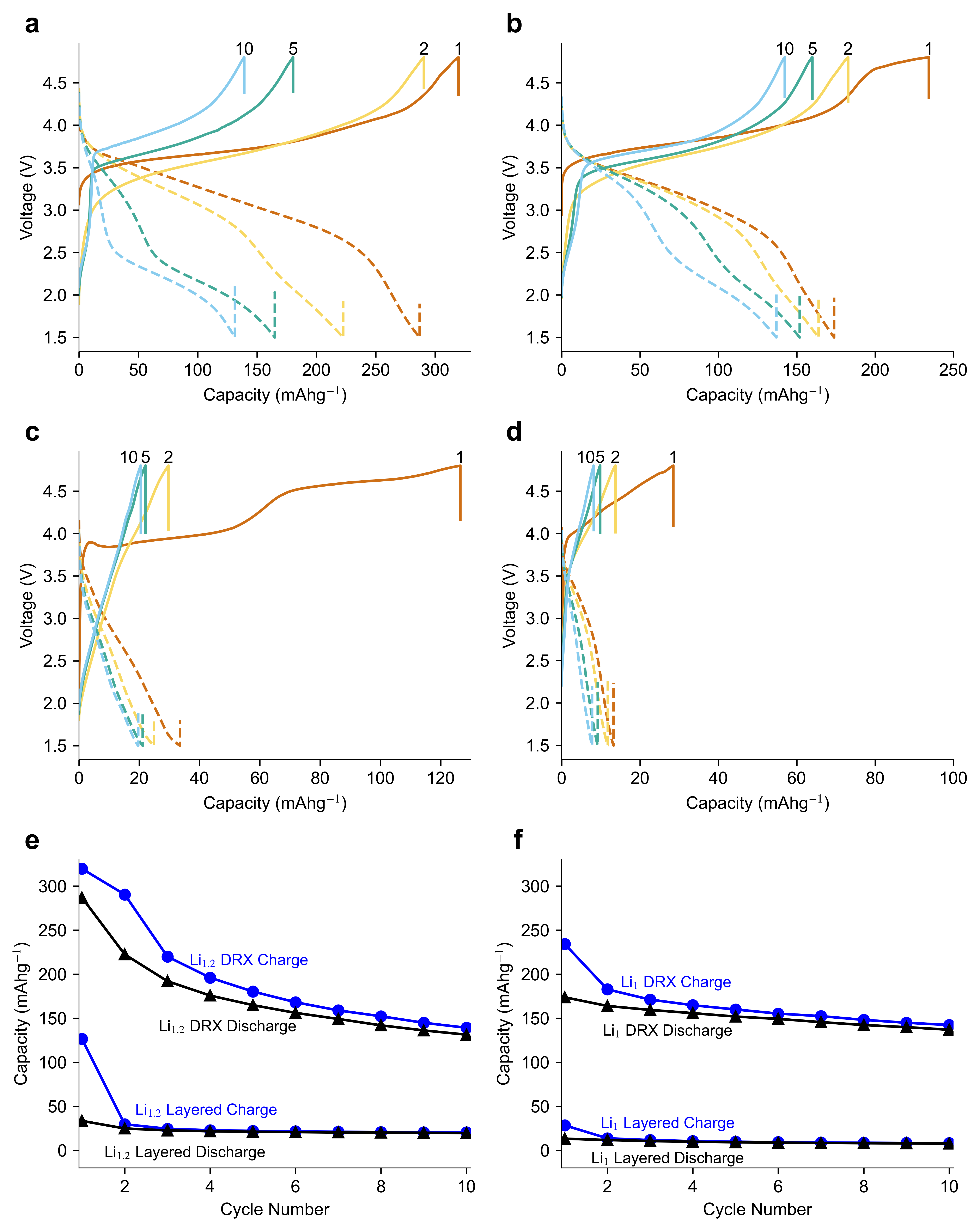}
    \caption{The charge/discharge curves between 1.5-4.8V at 0.1C rate for \textbf{a.} Li$_{1.2}$Cr$_{0.6}$Fe$_{0.2}$O$_2$ DRX \textbf{b.} LiCr$_{0.75}$Fe$_{0.25}$O$_2$ DRX \textbf{c.} Li$_{1.2}$Cr$_{0.6}$Fe$_{0.2}$O$_2$ Layered \textbf{d.} LiCr$_{0.75}$Fe$_{0.25}$O$_2$ Layered and the charge/discharge capacity for \textbf{e.} 20\% Li-excess \textbf{f.} non-Li-excess for both DRX and Layered structures cathodes. 
     }
    \label{fig:DRX_cycle}
\end{figure}

\begin{figure}
    \centering
    \includegraphics[width=0.8\textwidth]{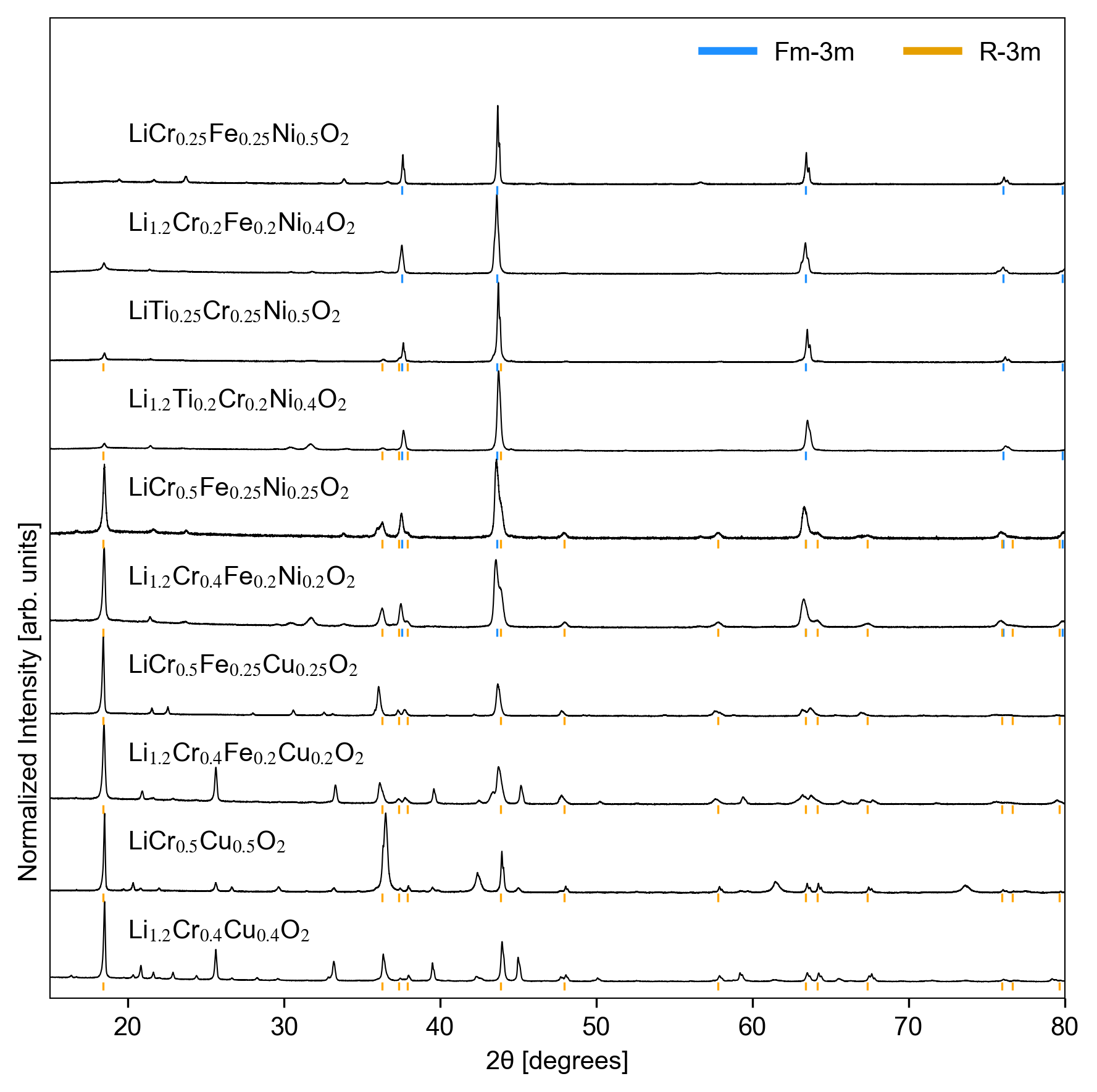}
    \caption{XRD spectrums of synthesized compounds and 20\% Li-excess variants. 
    Additional synthesis of selected compositions for their Li-excess variants were performed to demonstrate the similarity of the observed major phases (DRX and Layered) between no Li-excess vs. 20\% excess of Li, while the minor impurity phases might be different especially in the less stable $\mathrm{Li}\mathrm{Cr}_{0.5}\mathrm{Fe}_{0.25}\mathrm{Cu}_{0.25}\mathrm{O}_{2}$ and $\mathrm{Li}\mathrm{Cr}_{0.5}\mathrm{Cu}_{0.5}\mathrm{O}_{2}$. 
    We note that higher Li-excess levels beyond 20\% could potentially lead to additional changes in the synthesized phases. 
    Nevertheless, our work focuses on utilizing only the necessary Li-excess amount to facilitate Li diffusion, as further discussed in the SRO descriptor and cathode design sections.}
    \label{fig:DRX_Li12check}
\end{figure}

Given the strong LRO preference of Cr toward the Layered type, both LiCr$_{0.75}$Fe$_{0.25}$O$_2$ and Li$_{1.2}$Cr$_{0.6}$Fe$_{0.2}$O$_2$ are initially synthesized in Layered form as expected from F$_{\mathrm{Layered}}$ = -19 meV/atom at 1273 K, which, however, exhibit poor electrochemical performance initially. 
As predicted by F$_{\mathrm{DRX}}$ = 41.6 meV/atom, the DRX phase is far above the hull with high disordering temperature and so might be better accessed by non-equilibrium techniques. 
To utilize the favorable SRO tendency in its disordered phase, we performed additional ball-milling \cite{obrovac1998structure, sato2018metastable} to induce the transition from Layered to DRX phase, as shown in Figure~\ref{fig:DRX_CrFe12}.
Contrary to previous findings \cite{obrovac1998structure} in LiMO$_2$ with single $M$ = Co, Ni, Fe, and Ti, our LiCr$_{0.75}$Fe$_{0.25}$O$_2$ is first electrochemically inactive in its ordered Layered phases but enables much more Li diffusion and capacity after conversion into DRX, delivering an initial capacity of 234 mAhg$^{-1}$ and maintaining a reversible capacity of approximately 150 mAhg$^{-1}$ over 10 cycles without any Li-excess, as demonstrated in Figure~\ref{fig:DRX_cycle}b,f.
While the smaller particle size after ball-milling might contribute to an increased fraction of Li in the active diffusion network due to reduced transport length \cite{lee2021determining}, we propose that Li diffusion can be further facilitated by carefully selecting elements that favor Li$_4$ clustering.
Our results reinforce the relationship between the calculated SRO descriptor and experimentally measured active Li diffusion in DRX phases, which aligns with the positive trends discussed in the previous section and shown in Table \ref{tab:DRX_SRO}. 
These experimental results demonstrate that our design approach successfully allows us to select favorable element combinations for favorable Li-M local environments after disordering.

Even more favorable electrochemical behavior can be observed in the Li-excess variant \\Li$_{1.2}$Cr$_{0.6}$Fe$_{0.2}$O$_2$.
By combining excess Li and favorable SRO, the desired Li$_4$-rich percolating environment is further enhanced, boosting the accessible Li to 0.98 per formula unit with a high capacity of 320 mAhg$^{-1}$ in the initial charge. 
Although active Li diffusion is successfully achieved through our SRO design strategy, we note a low retention after cycling, especially in the Li-excess variant of Li-Cr-Fe-O cathodes. 
We hypothesize that this cycling behavior might result from oxygen loss, commonly observed in Li-excess Fe-containing cathodes \cite{yabuuchi2016origin, yamamoto2019charge}. 
Consistent with this idea, we note that the non-Li-excess LiCr$_{0.75}$Fe$_{0.25}$O$_2$ exhibits a much better retention rate, retaining even more capacity after 10 cycles compared to Li$_{1.2}$Cr$_{0.6}$Fe$_{0.2}$O$_2$.
The non-Li-excess structure features much fewer Li-O-Li configurations \cite{seo2016structural}, likely resulting in less oxygen redox activity and therefore less degradation.
Other potential causes of capacity fade include the decomposition of the electrolyte at high voltage \cite{liu2021hydrolysis}, dissolution of TM \cite{geng2021operando, zhan2018dissolution}, and various mechanisms leading to structural instability after cycling.
These kinetic mechanisms were not specifically part of our ordering design framework (addressing Li diffusion and phase stability through ordering design in pristine materials) and should be considered separately.
As far as future cathode design, in addition to the conventional Ni, Mn, Co, and Al as element choices, compositions with Cr, Fe, V, Ti, and Mg might be feasible and practical considering both Li diffusion and phase stability as well as their comparatively lower cost. For specialized applications or research purposes due to their high cost, Ga or Rh could also be considered. 

\section{Conclusion}

In this work, we conducted a large first-principles study of elemental ordering tendencies in Li metal oxides, identifying ordering tendencies that correspond with favorable rocksalt phase stability and the likelihood of active Li diffusion. 
We demonstrated a practical approach to study ordering tendencies by leveraging a comprehensive first-principles computational database supported by a selected set of experimental data. 
The computational framework enables the rational selection of elements, opening new pathways for customizing ordering in battery cathode materials design. 
In a proof-of-concept case study on the Li-Cr-Fe-O system, we achieved a substantial 234 mAhg$^{-1}$ initial charge capacity in the non-Li-excess and 320 mAhg$^{-1}$ in 20\% Li-excess DRX cathode by applying post-processing ball millings.
A similar development strategy can be applied to thousands of new battery chemistries, potentially accelerating the discovery of high-capacity cathodes with improved stability and Li diffusion.


\begin{acknowledgement}

The authors thank Joseph Montoya and Abraham Anapolsky for their helpful discussions. T.-c.L., B.B., A.S.-C., and Y.Z. were supported by funding from the Toyota Research Institute. T.-c.L. acknowledges funding from the Taiwanese government fellowship for Overseas study. T.-c.L. thanks Shima Shahabfar for the support with the crystal structure visualization upgrades in Figures 1a and 3a. A.S.-C. acknowledges the financial support to the National Agency for Research and Development (ANID)/DOCTORADO BECAS CHILE/2018 - 56180024. The authors acknowledge computational resources from the Quest high-performance computing facility at Northwestern University which is jointly supported by the Office of the Provost, the Office for Research, and Northwestern University Information Technology, and also the computational resources from the National Energy Research Scientific Computing Center (NERSC), a U.S. Department of Energy Office of Science User Facility located at Lawrence Berkeley National Laboratory, operated under Contract No. DE-AC02-05CH11231 using NERSC award BES-ERCAP23792, to perform DFT calculations.  
\end{acknowledgement}

\section*{Competing Interests}
U.S. Pat. App. No. 18/897,372, U.S. Pat. App. No. 18/897,588, and U.S. Pat. App. No. 63/603,828 have been jointly filed by Toyota Research Institute, Toyota Motor Company, and Northwestern University, reflecting work in this manuscript with all authors listed as inventors. 
The authors are planning to file further patent applications on the results of this work.

\section*{Correspondence}
Correspondence and requests for computational materials should be addressed to Chris Wolverton. Those related to experimental materials should be addressed to Steven Torrisi.

\section*{Data Availability Statement}
The data used to produce this manuscript is available upon reasonable request. The full database of compositions and first-principles calculations will be released in the Open Quantum Materials Database and/or a separate article after publication of this current work.  

\section*{Technology Usage Statement}
Grammar and fluency checks were performed on the earlier draft through the responsible use of AI tools such as ChatGPT and Grammarly Premium, with generated suggestions considered but not necessarily applied, followed by substantial revisions from authors.  
All authors have reviewed and approved all information demonstrated in this work.  


\bibliography{reference}

\clearpage

\section*{Supporting Information}

\begin{figure}
    \centering
    \includegraphics[width=0.77\textwidth]{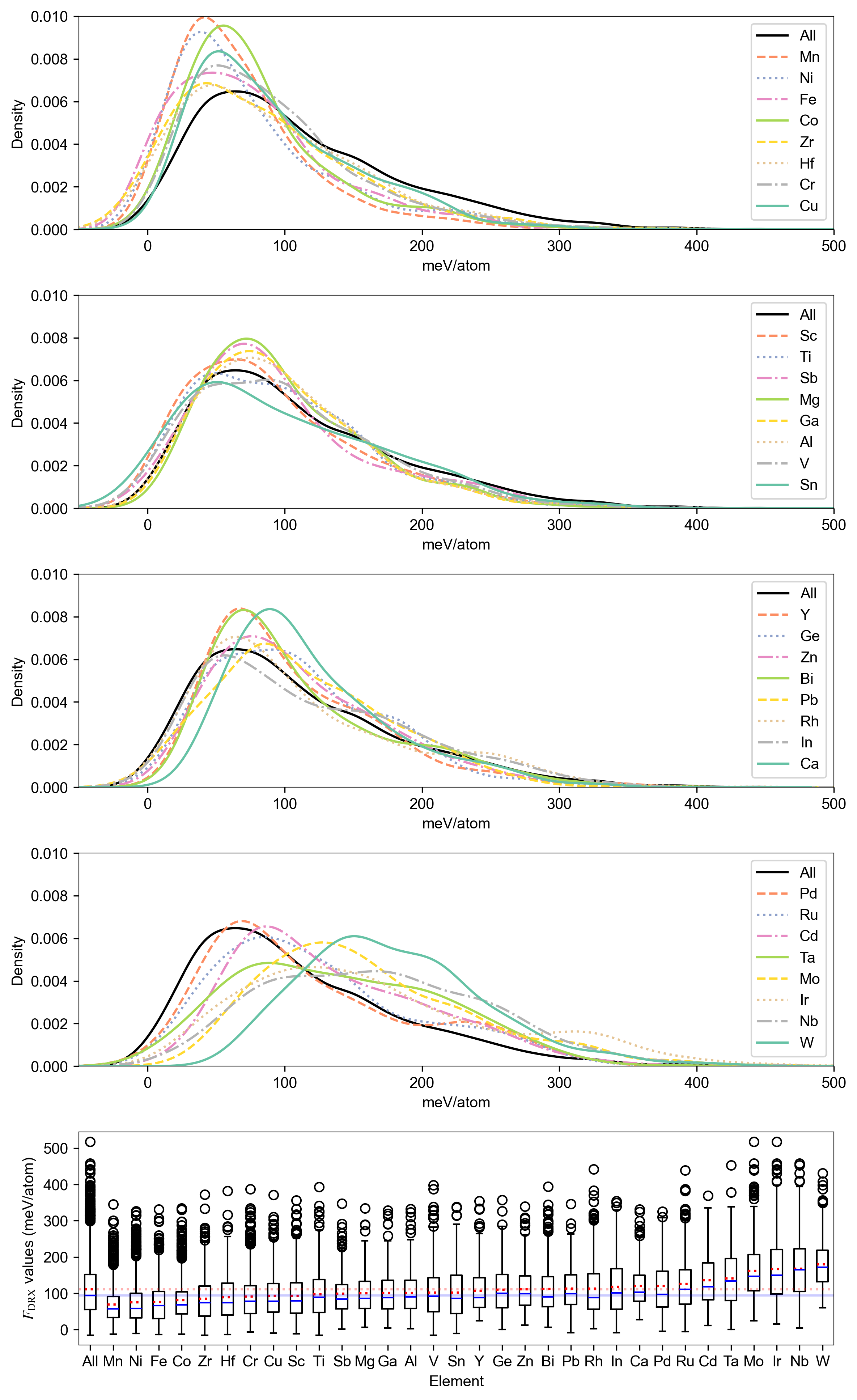}
    \caption{Phase stability descriptors distributions for LiMO$_2$ DRX phase. Distributions of F$_{\mathrm{DRX}}$ from the entire computational database.  (Continued on next page.)} 
    \label{fig:FDRX_stat}
\end{figure}
\begin{figure}
  \ContinuedFloat
  \caption{(cont.) The black curve plotted in each subplot, labeled “All,” represents the F$_{\mathrm{DRX}}$ distribution of all compositions in the entire database.
  Individual curves for each element depict distributions calculated from all compositions containing that specific element. 
  The last subplot is the box plot that indicates the range and key values of the F$_{\mathrm{DRX}}$ distribution, including medians (solid blue lines) and means (red dot lines). The horizontal lines plotted in the background of the box plot are from “All” compositions for comparisons. 
  Elements are sorted by their mean values. 
  We noted that, while elements like Cr, Cu, Sb, Mg, Ga, Al, and Y exhibit lower median values than “All”, only a few compositions with these elements at the lower end can pass the stable DRX threshold due to smaller standard deviations in distributions, which excludes them from being listed as beneficial elements for DRX in Figure~\ref{fig:DRX_PSD}c.
    }
\end{figure}

\clearpage

\begin{figure}
    \centering
    \includegraphics[width=0.77\textwidth]{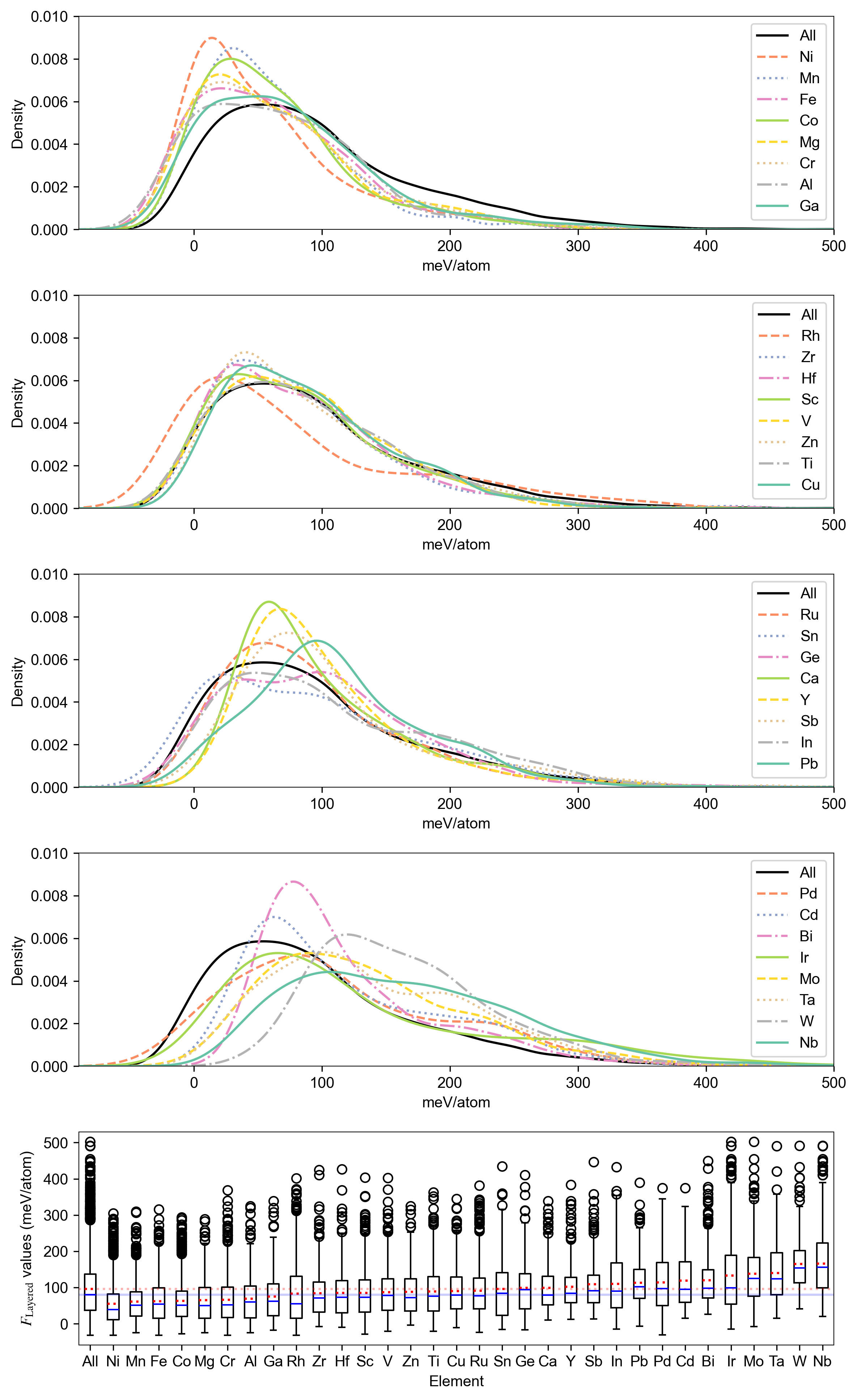}
    \caption{Phase stability descriptors distributions for LiMO$_2$ Layered phase. Distributions of F$_{\mathrm{Layered}}$ from the entire computational database. (Continued on next page.)} 
    \label{fig:FLayered_stat}
\end{figure}
\begin{figure}
  \ContinuedFloat
  \caption{(cont.) The black curve plotted in each subplot, labeled “All,” represents the F$_{\mathrm{Layered}}$ distribution of all compositions in the entire database.
  Individual curves for each element depict distributions calculated from all compositions containing that specific element. 
  The last subplot is the box plot that indicates the range and key values of the F$_{\mathrm{Layered}}$ distribution, including medians (solid blue lines) and means (red dot lines). The horizontal lines plotted in the background of the box plot are from “All” compositions for comparisons. 
  Elements are sorted by their mean values. 
  Unlike statistics of F$_{\mathrm{DRX}}$ distributions, elements with lower median F$_{\mathrm{Layered}}$ are consistent with favorable elements for Layered phase in Figure~\ref{fig:DRX_PSD}c.
    }
\end{figure}

\clearpage
\begin{figure}
    \centering
    \includegraphics[width=0.77\textwidth]{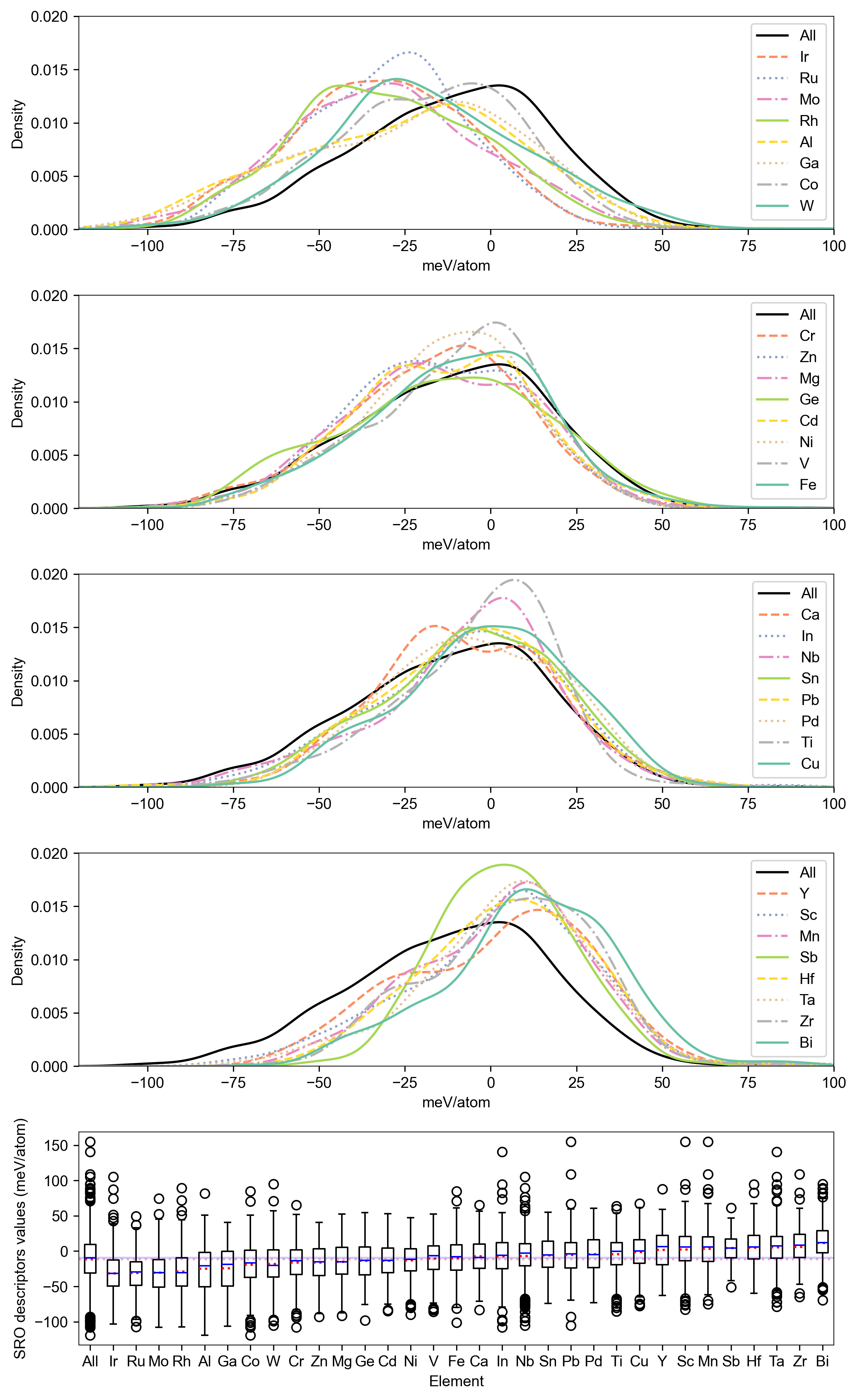}
    \caption{SRO descriptors distributions. Distributions of SRO descriptors from the entire computational database. (Continued on next page.)} 
    \label{fig:SRO_stat}
\end{figure}
\begin{figure}
  \ContinuedFloat
  \caption{(cont.) The black curve plotted in each subplot, labeled “All,” represents the SRO distribution of all compositions in the entire database.
  Individual curves for each element depict distributions calculated from all compositions containing that specific element. 
  The last subplot is the box plot that indicates the range and key values of the SRO descriptor distribution, including medians (solid blue lines) and means (red dot lines). 
  The horizontal lines plotted in the background of the box plot are from “All” compositions for comparisons. 
  Elements are sorted by their mean values, which is in the sequence different from but similar to the sequence in Figure~\ref{fig:DRX_SROD}c of the main text (sorted by the percentage of Li-M mixing compositions.)
    }
\end{figure}

\clearpage
\subsection{Supporting Note 1 : Plotting distribution curves }
Distribution curves (in Figure~ \ref{fig:DRX_PSD}a,b, Figure~\ref{fig:DRX_SROD}b, Figure~ \ref{fig:FDRX_stat},\ref{fig:FLayered_stat},\ref{fig:SRO_stat}) are plotted with kernel density estimations using the implementation in the pandas library 1.4.3, namely pandas.DataFrame .plot(kind=’kde’). However, the reported percentages in Figure~\ref{fig:DRX_PSD} and Figure~~\ref{fig:DRX_SROD} are calculated using the precise number of compositions in each class from the computational database instead of the area under curves.  

\subsection{Supporting Note 2 : Elemental ordering statistics is not sensitive to the threshold}
\subsubsection{Phase stability descriptors threshold}
In the elemental statistics section for rocksalt phase stability, we primarily focused on the left tail of distributions that could pass (value being lower than) the empirically determined threshold values as supported by experimental synthesis rather than on the mean or median of the entire distributions.
In Figure~\ref{fig:FDRX_stat} and \ref{fig:FLayered_stat}, we can observe that most distributions show monotonic characteristics around the threshold values. This simplicity is beneficial for a ranking-style analysis across elements with the same numerical thresholds, making the elemental statistics less sensitive to uncertainties in threshold values (which can be mitigated by conducting more experiments) and to uncertainties arising from the ideal mixing approximations made in computational free energy values.
More straightforwardly, we show in the following Figure~\ref{fig:SI_F_sens} a series of periodic-table-style heat maps (similar to Figure~\ref{fig:DRX_PSD}c, original threshold values: 7 meV/atom for DRX and -11 meV/atom for Layered phase) of changing threshold values across the range of -10 to +20 meV/atom from the original values. 
Decreasing the value by more than -10 meV/atom would result in very few compositions passing the shifted threshold. 
Increasing the threshold values allows more compositions to be classified as “stable,” as seen in the increasing stable DRX/Layered percentages, but comparison between elements, as demonstrated by the colors, are nearly unchanged.
This comparison validates the reliability of the large-scale elemental ordering stability investigation presented in the main text. 
Nevertheless, the empirically defined thresholds, supported by our experimental results, still offer better physical robustness.   

\begin{figure}
    \centering
    \includegraphics[width=0.75\textwidth]{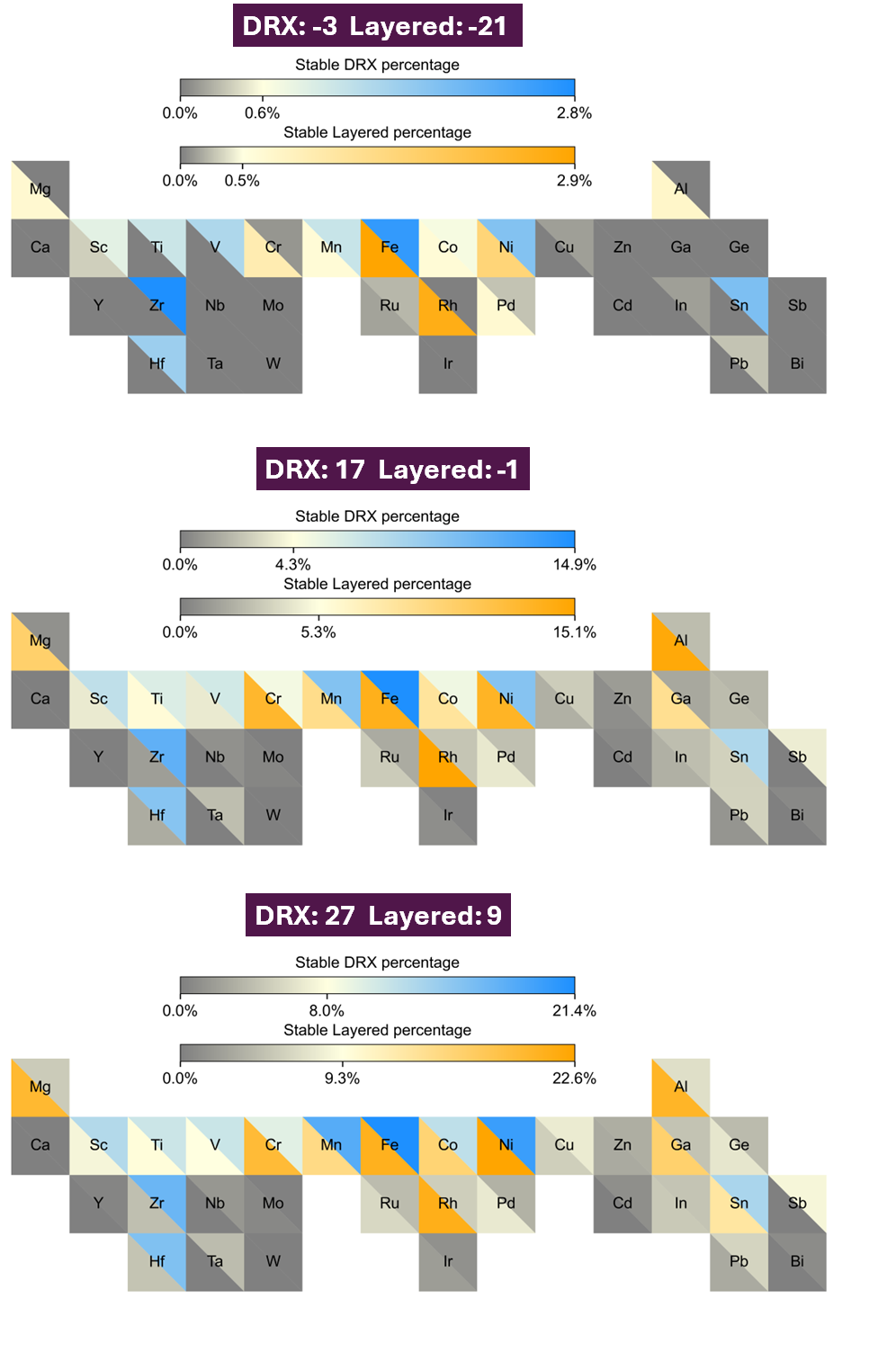}
    \caption{A series of Periodic-table-style heat maps for the stable DRX and Layered percentages but with changing thresholds on both phases: -10, +10, and +20 meV/atom from the experimentally defined threshold (DRX: 7 meV/atom, Layered: -11 meV/atom). Numbers in the purple box are the F$_{\mathrm{DRX}}$ and F$_{\mathrm{Layered}}$ threshold values (in meV/atom) utilized to generate each plot. } 
    \label{fig:SI_F_sens}
\end{figure}
\clearpage
\subsubsection{SRO descriptors threshold}
\begin{figure}
    \centering
    \includegraphics[width=0.9\textwidth]{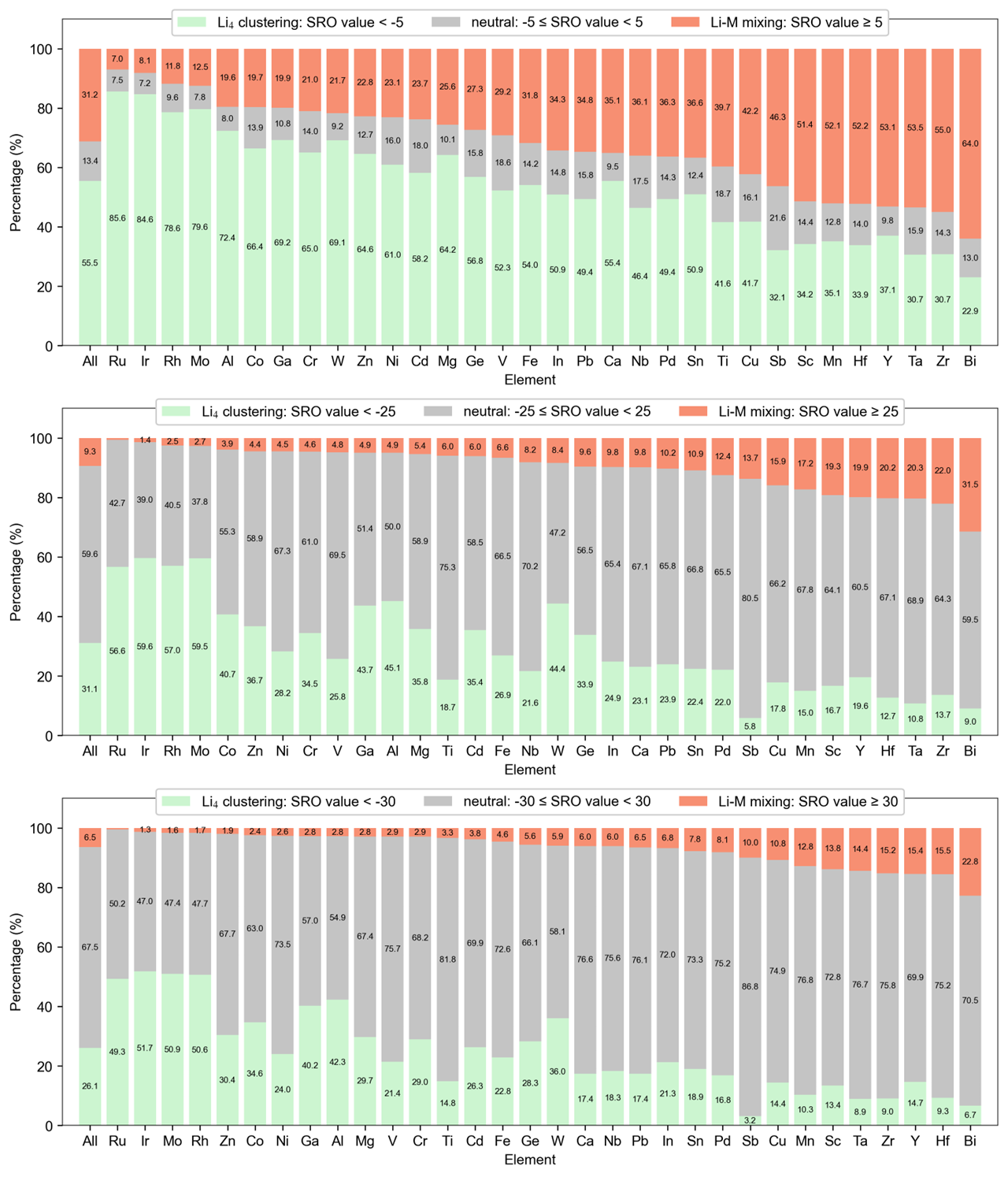}
    \caption{A series of plots similar to Figure~\ref{fig:DRX_SROD}c (originally using a threshold magnitude of 15 meV/atom) is presented, but with varying SRO classification thresholds magnitudes (5, 25, and 30 meV/atom).  
    Although altering the threshold changes the percentage of compositions in each category, the elemental rankings remain similar, as quantified by the Spearman’s rank correlation in the corresponding discussion.} 
    \label{fig:SI_SRO}
\end{figure}

We conducted a similar sensitivity test on the SRO descriptor. In Figure~\ref{fig:DRX_SROD}c, elements are sorted by the percentage of compositions classified under “Li-M mixing” category to avoid interruptions in Li diffusion channels. 
In Figure~\ref{fig:SRO_stat}, “Li-M mixing” corresponds to the right tail of the distributions, exhibiting monotonic characteristics around the thresholds. 
We show in Figure~\ref{fig:SI_SRO} a series of plots similar to Figure~\ref{fig:DRX_SROD}c of changing both positve and negative threshold values (boundary of ``neutral'') across the range of 5 to 30 meV/atom. 
While the percentages in categories can vary drastically, the overall elemental ranking remains largely unaffected by threshold variations, as further quantified by the Spearman’s rank correlation of the elemental rankings in Table~\ref{Spearman}.\\

\begin{table}
  \centering
  \caption{Spearman’s rank correlation matrix of the elemental rankings from varying threshold magnitudes (meV/atom).}
  {
\renewcommand{\arraystretch}{1.2}
\begin{tabular}{c|cccccc}
  \hline\hline
  Thresholds & 5 & 10 & 15 & 20 & 25 & 30 \\ \hline
   5  & 1.000 & 0.983 & 0.971 & 0.954 & 0.938 & 0.941 \\
  10  & 0.983 & 1.000 & 0.985 & 0.975 & 0.964 & 0.965 \\
  15  & 0.971 & 0.985 & 1.000 & 0.993 & 0.982 & 0.980 \\
  20  & 0.954 & 0.975 & 0.993 & 1.000 & 0.985 & 0.986 \\
  25  & 0.938 & 0.964 & 0.982 & 0.985 & 1.000 & 0.986 \\
  30  & 0.941 & 0.965 & 0.980 & 0.986 & 0.986 & 1.000 \\ \hline\hline
\end{tabular}

  }\label{Spearman}
\end{table}

\clearpage

\subsection{Supporting Note 3 : Justification of fine relaxations on synthesized compositions}
Given the complex nature of the low-symmetry supercells, we anticipated that the structures relaxed by DFT calculations using the default OQMD convergence criteria might not be fully optimized to reach the global minimum of the complex potential surface. 
As mentioned in the method section for DFT calculations, extra ionic fine relaxations were performed on 36 supercells of 9 synthesized compositions with 10$^{-2}$ eV/Å force convergence criteria, generally leading to structures in a more relaxed arrangement at the cost of increased ionic steps and resource consumption.
These further optimized structures undergo final high-fidelity static calculations identical to the OQMD default workflow. 
Due to resource constraints, fine relaxations are currently performed only for synthesized compositions, which are utilized to define the critical threshold for stable composition classification.
Since the E$_{\mathrm{hull}}$ is calculated by the energy difference of the supercell and the ground states, which are computed with OQMD default convergence settings, the discrepancy in the convergence settings may raise concerns about comparing structures with inconsistent settings. 
Nevertheless, we claim that ground states on the convex hull are typically simple and can be easily relaxed to the global minimum. 
We anticipate that comparing these further optimized structures of supercells with the ground states provides a more accurate physical insight into the stability of complex supercells by eliminating the residual forces among atoms. 
Adhering to this concept, we only adjust the final E$_{\mathrm{hull}}$ if the energy from the final static calculation is lower than the value obtained from the OQMD default workflow, thereby considering the new structure more optimized. 
The variations in energy after additional fine relaxations are documented in Table \ref{Ediff}, where we noted that 2 out of 36 structures did not show improvement even after fine relaxations; hence, their E$_{\mathrm{hull}}$ values remain unchanged. 
The energy scale of these numbers within the table can provide a rough estimate of the uncertainty in the DFT calculations for complex supercells across the entire database. In the table, the average change is -6.6 meV/atom, with only three fine calculations showing a change greater than -10 meV/atom.\\

\begin{table}
  \centering
  \caption{Energy changes after fine structural relaxations (meV/atom).}
  {
  \renewcommand{\arraystretch}{1.2}
  \begin{tabular}{c r r r r}
    \hline
    Composition & DRX & Layered & Spinel-like & $\gamma$-LiFeO$_2$ \\
    \hline
    Li$_4$CrFeNi$_2$O$_8$   &  $-6.9$ &  $-6.5$ &  $-6.8$ &  $-6.8$ \\ \hline
    Li$_4$CrFe$_2$NiO$_8$   &  $-6.1$ &  $-4.3$ &  $-4.1$ &  $-6.8$ \\ \hline
    Li$_4$Cr$_2$FeNiO$_8$   &   0.0   & $-18.7$ &  $-3.8$ &  $-6.5$ \\ \hline
    Li$_4$TiCrNi$_2$O$_8$   &  $-4.5$ &  $-3.9$ &  $-6.3$ &  $-6.9$ \\ \hline
    Li$_4$Cr$_2$FeCuO$_8$   &  $-5.5$ &   0.0   &  $-8.2$ &  $-6.7$ \\ \hline
    Li$_2$CrCuO$_4$         &  $-7.7$ &  $-3.3$ & $-15.1$ &  $-8.4$ \\ \hline
    Li$_4$Cr$_2$GaFeO$_8$   &  $-4.8$ &  $-4.4$ &  $-4.0$ &  $-6.5$ \\ \hline
    Li$_4$Cr$_3$FeO$_8$     &  $-9.4$ &  $-4.1$ &  $-4.0$ &  $-6.3$ \\ \hline
    Li$_4$TiCr$_2$CuO$_8$   &  $-7.2$ &  $-4.9$ &  $-7.0$ & $-21.8$ \\ \hline
  \end{tabular}
  }
  \label{Ediff}
\end{table}

\end{document}